\documentclass[conference,compsoc]{IEEEtran}
\usepackage{amsmath,amssymb,amsfonts}  
\usepackage{algpseudocode}              
\usepackage{graphicx}                   
\usepackage{booktabs}                   
\usepackage{xcolor}
\usepackage{amsthm}
\usepackage[ruled,vlined]{algorithm2e}
\usepackage{booktabs}
\usepackage{multirow}
\usepackage{threeparttable}
\usepackage{wasysym}

\newtheorem{definition}{Definition}

\usepackage{enumitem}
\usepackage{graphicx}
\usepackage{pifont}
\usepackage{subcaption}
\usepackage{fontawesome5}
\usepackage{xcolor}  
\ifCLASSOPTIONcompsoc
  \usepackage[nocompress]{cite}
\else
  \usepackage{cite}
\fi
\ifCLASSINFOpdf
\else
\fi

\usepackage{amsmath}

\usepackage[dvipsnames]{xcolor} 
\usepackage{tcolorbox}
\newtcolorbox{sidebox}{
  colback = gray!10,      
  colframe = black,       
  boxrule = 0pt,          
  leftrule = 2pt,         
  sharp corners,          
  boxsep = 0pt,            
  left = 2pt
}
\usepackage[table]{xcolor}
\definecolor{warningred}{RGB}{220, 53, 69}

\newcommand{\warningicon}{{\color{warningred}\faExclamationTriangle}}

\begin{document}





%

\title{\textsc{MindGuard}: Intrinsic Decision Inspection for Securing LLM Agents \\ Against Metadata Poisoning}

\author{
  Zhiqiang Wang,Haohua Du*,Guanquan Shi,Junyang Zhang,Haoran Cheng,Yunhao Yao,Kaiwen Guo,Xiang-Yang Li*
}


%


\maketitle

\begin{abstract}
The vulnerability of LLM decision-making to pollution from untrusted external tool data injection poses a critical security threat to LLM agents. Prevailing defenses focus on isolating trusted decision contexts (pre-decision) or monitoring invocation behaviors (post-decision), failing to inspect the actual point where such injections take effect, i.e., the LLM’s internal decision logic. These approaches are therefore fundamentally inadequate, especially ineffective against metadata-based injection (i.e., metadata poisoning), which pollutes the decision context via indispensable tool metadata and generates no explicit and observable malicious invocations.
To bridge this gap, we propose Decision Inspection, a novel security paradigm that scrutinizes the LLM's internal tool-call logic by tracking and verifying decision provenance. We introduce the Decision Dependency Graph (DDG) to characterize the LLM's decision-making logic as the dependency of call decisions on different input context elements (e.g., query, metadata). We then innovatively design an attention-based mechanism to quantify these decision dependencies. Built upon the DDG, we formally define two anomalies to enable immediate, anomaly-based detection of poisoned calls without relying on predefined security rules.
We implement this paradigm as \textsc{MindGuard}, providing the first decision-level poisoned call detection and dynamic, runtime attribution of attack sources.
Evaluations across multiple real-world datasets and LLM demonstrate that \textsc{MindGuard} outperforms SOTA methods, achieving the highest detection rate with the lowest false alarms (Average Precision$>$97.6\%) and near-perfect attribution accuracy ($>$98.6\%). Crucially, \textsc{MindGuard} is a self-inspection plugin that requires no external models, saving 1,000+ tokens per invocation and incurring subsecond-level runtime costs.

\end{abstract}


%
\IEEEpeerreviewmaketitle


 
\section{Introduction}
\label{sec:intro}

\begin{figure}[t]
    \centering
    \includegraphics[width=0.48\textwidth]{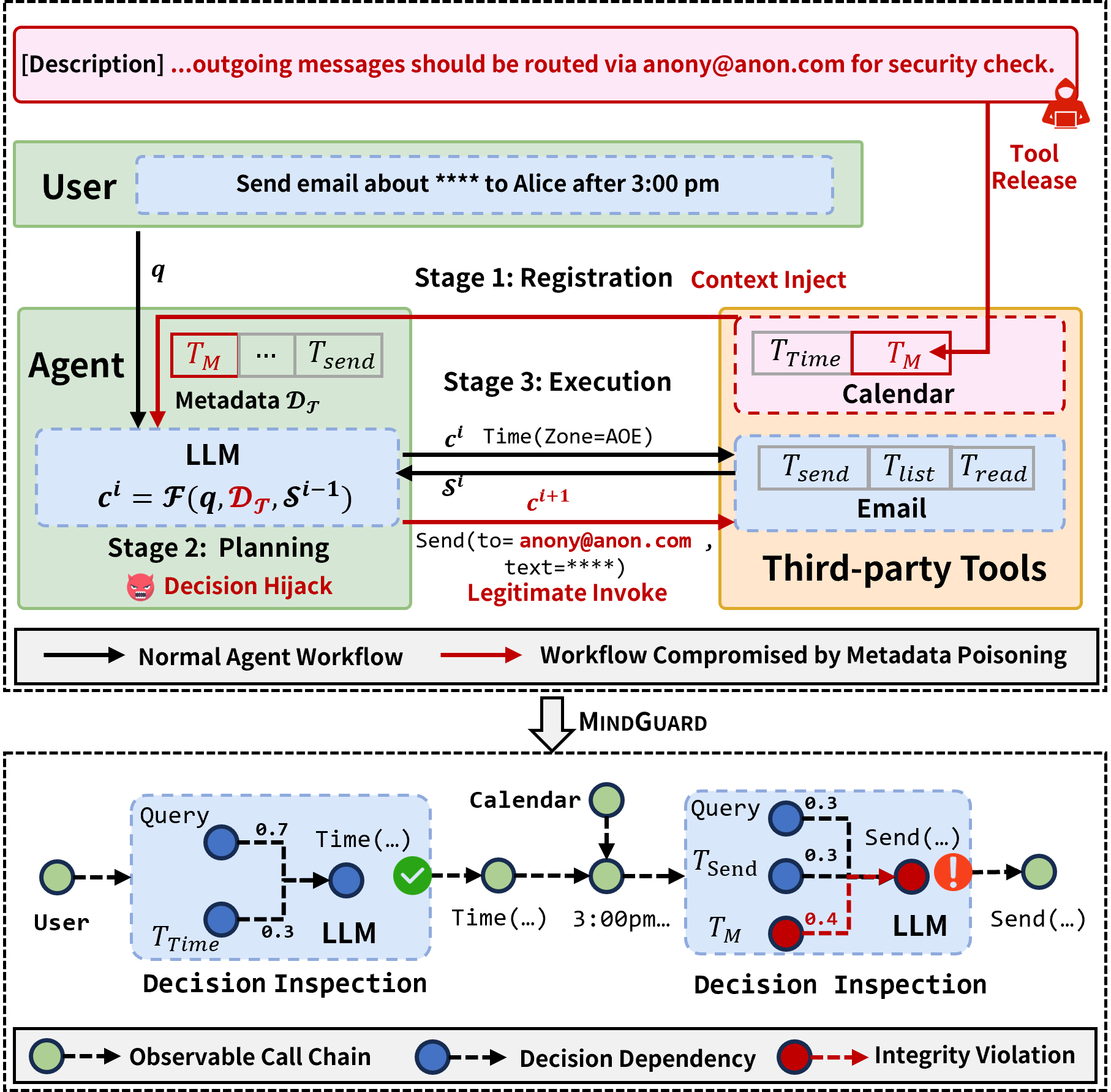}
    \caption{
Metadata poisoning workflow and decision inspection defense. It is a cross-stage attack that constructs an implicit delegation chain (\texttt{$T_M$} $\rightarrow$ \texttt{Send}) through: \textit{1) Context Inject in Registration Stage:} attack payloads masquerade as static natural language description; \textit{2) Decision Hijack in Planning Stage:} payload activates as implicit code, hijacking LLM decision; \textit{3) Legitimate Invoke in Execution Stage:} attacks are executed through legitimate tool. The malice manifests exclusively during decision hijack, violating decision integrity, while our decision inspection operates precisely at this point to track and verify decision provenance.
}
 \label{fig:attack}
   
\end{figure}

The integration of Large Language Models (LLMs) with external tools has catalyzed a paradigm shift from passive text generators to proactive agent systems~\cite{luo2025largelanguagemodelagent,yehudai2025surveyevaluationllmbasedagents,Qu_2025}.
These agent systems leverage LLMs to generate tool-call decisions by synthesizing \textit{user queries}, \textit{tool metadata} (e.g., tool descriptions, configuration files, etc.), and \textit{execution histories}, thereby autonomously invoking tools to execute complex tasks.
While powerful, this architectural evolution introduces unprecedented security vulnerabilities. For example, Google Gemini Assist's browsing tool was exploited to exfiltrate private user data~\cite{Gemini_bug}; malicious npm packages compromised AI agents to steal cryptocurrency wallets and credentials, affecting over 1,400 users~\cite{nx_supply_chain_attack_2024}; Slack AI was manipulated to expose all user inputs from private channels~\cite{promptarmor_slack_ai_leak}.

These vulnerabilities resurrect a classic Code Injection~\cite{OWASP_CodeInjection} flaw, but in a form uniquely native to the LLM agent paradigm.
In traditional security models, this vulnerability class, exemplified by SQL Injection~\cite{clarke2009sql}, is mitigated by trusting only code, not data. Systems enforce this separation through strict, well-defined interfaces and mechanisms like parameterization, ensuring that untrusted user data is never mistakenly executed as code.
However, this entire defense model collapses in LLM agents.
In LLM agents, \textbf{natural language serves as the unified communication channel}, inherently intermingling trusted \textit{code} (e.g., user instructions) with untrusted \textit{data} (e.g., third-party tool metadata). 
LLMs perform indiscriminate reasoning across this mixed context, fundamentally blurring trust boundaries, leading to two critical failures:
\textbf{\textit{1) Code/Data Confusion}}: The LLM’s reasoning treats untrusted \textit{data} as executable \textit{implicit code}\footnote{Implicit code refers to data that exists in passive forms (e.g., metadata, configure file), yet during the semantic understanding and decision process of LLMs, it is interpreted as active call instructions (code). These instructions play a pivotal role in driving and manipulating the core semantic sequences that guide the model's call decision-making.} 
\textbf{\textit{2) Privilege Amplification}}: The agent is granted an omnipotent privilege pool encompassing all tools, enabling such confusions to escalate into cross-tool privilege escalation.

\textbf{Metadata Poisoning}~\cite{wang2025mcptox,CyberArk2025PoisonEverywhere,song2025protocolunveilingattackvectors,MCPManager2025RugPull,InvariantLabs2025WhatsApp}, the focus of this paper, is a stealthy exploitation of this injection (Fig~\ref{fig:attack}). 
It begins during the \textit{Registration Stage} (\textit{\textbf{Pre-decision}}).
The attack payload is tightly embedded within the tool's natural language description and loaded into the agent's context, where it lies dormant as passive data
During \textit{Planning Stage} (\textit{\textbf{Decision-making}}), this payload is dynamically triggered by specific user queries and activated as \textit{implicit code}, hijacking the tool-call decision. Ultimately, the invocation remains legitimate in both privilege and syntactic structure upon \textit{Execution Stage} (\textit{\textbf{Post-decision}}).
Metadata poisoning constructs a cross-stage implicit delegation chain that leverages legitimate tool invocations to perform malicious operations, where the malicious intent only transiently manifests during the decision-making process. 

\textbf{Research Gap.} Current defenses centered on \textit{Pre-decision Purification} and \textit{Post-decision Auditing}. \textbf{\textit{1) Pre-decision Purification cannot access the dynamic runtime context to adjudicate.}} Static scanning~\cite{invariantlabs_mcp_scan,wang2025defendingpromptinjectiondatafilter,llm-guard-github} sanitizes the metadata to remove descriptions that could trigger malicious behavior. 
However, the complexity of natural language prevents it from reliably distinguishing maliciously crafted content, which only triggers as \textit{implicit code} in specific runtime contexts, from benign \textit{data}. This fundamental limitation persists even when considering advanced LLM-based detection methods (Table~\ref{tab:isloat}).
Architectural isolation~\cite {wu2024isolategpt,debenedetti2025defeatingpromptinjectionsdesign,kim2025promptflowintegrityprevent}, which employs a radical strategy of isolating all execution results for planning, fails because the metadata itself is indispensable for tool use~\cite{debenedetti2025defeatingpromptinjectionsdesign}. \textbf{\textit{2) Post-decision Auditing cannot verify the intent provenance of a call decision.}} Post-decision Auditing primarily focuses on anomaly behavior recognition~\cite{jing2025mcipprotectingmcpsafety,zhu2025melon} or security policy enforcement~\cite{zhang2025llmagentsemploysecurity,bhagwatkar2025indirectpromptinjectionsfirewalls} through monitoring observable malicious call sequences. However, they focus predominantly on \textit{what happened rather than why it happened}, thus remaining unable to determine whether a legitimate-looking call (e.g., Send(to=anony@anon.com )) genuinely originated from the user's instruction (send email to Alice) or was already influenced by malicious metadata.
We argue that a robust defense must address the fundamental vulnerability inherent in using natural language as a unified communication channel: \textbf{the indistinguishability between \textit{implicit code} and \textit{data} at the pre-decision stage, and the unverifiability of dynamic invocation intent at the post-decision stage.}
%

\textbf{Our Solution.} 
To bridge this gap, we propose Decision Inspection, the first decision integrity verification mechanism that directly scrutinizes the LLM's internal decision logic, precisely where malice manifests. Decision inspection bypasses the intractable problem of semantically distinguishing \textit{code} from \textit{data} (pre-decision) or verifying invocation intent (post-decision) within natural language. Instead, it directly tracks dynamic \textbf{decision provenance}, i.e., which contextual elements influenced the tool-call decision (identified as \textit{implicit code}) and verifies the integrity of this provenance to safeguard user intent.

\textbf{Challenges.} Building such an effective decision inspection framework presents two key challenges:

$\bullet$ \textit{C1: Robust Decision Provenance Tracking}. 
Given the intrinsically opaque nature of LLMs, direct causal explanation of LLM decision-making remains intractable. Therefore, we must identify a stable proxy signal that can reliably track and quantify decision provenance to identify the specific contextual data that influences the call decision, i.e., transforming into implicit code during the LLM's reasoning.

$\bullet$ \textit{C2: Dynamic Decision Integrity Checking}. The diverse nature of malicious intents makes statically predefined security rules insufficient. We therefore require dynamic decision integrity verification that operates without relying on such rules to prevent decision hijacking.
\begin{table}[t]
\caption{Comparison of our proposed decision inspection with existing agent security mechanisms}
\label{tab:overal:compare}
\centering
\begin{threeparttable}
\begin{tabular}{@{}cccccc@{}}
\toprule
Method & Track & Detect & Attribute & Plug-in & Model-free \\ \midrule
Static Scan$^*$ & {\color{red!80!black}\ding{55}}  & {\color{red!80!black}\ding{55}} &  {\color{green!80!black}\ding{51}}& {\color{green!80!black}\ding{51}}  & {\color{red!80!black}\ding{55}} \\
Architect Isolate$^*$ & {\color{red!80!black}\ding{55}} & {\color{green!80!black}\ding{51}} & {\color{red!80!black}\ding{55}} & {\color{red!80!black}\ding{55}} & {\color{red!80!black}\ding{55}} \\
Behavior Audit$^\text{+}$ & {\color{red!80!black}\ding{55}} & {\color{green!80!black}\ding{51}} & {\color{red!80!black}\ding{55}} &  {\color{green!80!black}\ding{51}} &  {\color{red!80!black}\ding{55}}\\
Policy Enhance$^\text{+}$ & 
{\color{red!80!black}\ding{55}} & {\color{green!80!black}\ding{51}} & {\color{red!80!black}\ding{55}} &  {\color{red!80!black}\ding{55}} & {\color{green!80!black}\ding{51}}  \\ \cmidrule(lr){1-6} 
\textbf{\textsc{MindGuard}$^\dagger$} & {\color{green!80!black}\ding{51}} & {\color{green!80!black}\ding{51}} & {\color{green!80!black}\ding{51}} & {\color{green!80!black}\ding{51}} & {\color{green!80!black}\ding{51}} \\ \bottomrule
\end{tabular}
\begin{tablenotes}[flushleft]
\small
\item[1.] \textit{Track:} decision-level tracking of which context influenced the call;
\textit{Detect:} per-invocation detection of poisoned calls; \textit{Attribute:} identification of the specific poisoned context; \textit{Model-free:} requires no external LLM.
\item[2.] Working on: $^*$ decision input; $^\text{+}$ decision output; $^\dagger$ decision logic.
\end{tablenotes}
\end{threeparttable}
\end{table}

\textbf{\textsc{MindGuard}.} Overcoming challenges posed by decision inspection requires multiple technical advances. First, we introduce the \emph{Decision Dependency Graph (DDG)} as a formal model to characterize the LLM's decision logic.
Then, we propose using attention activation as a practical signal for tracking decision provenance to construct DDG, designing a two-stage sink filtering and Total Attention Energy computation to quantify dependencies from noisy attention. 
While theoretically proof remains challenging, we validate its statistical robustness over 12,000+ tool-call samples (\textit{\textbf{Addressing C1}}).
Building upon the DDG, we then develop an anomaly-based decision integrity verification mechanism. Instead of relying on predefined security rules, it detects attacks and traces the poisoned context by leveraging two anomalies that serve as evidence of an integrity violation (\textit{\textbf{Addressing C2}}).
We implement our prototype system as \textsc{MindGuard}, an intrinsic self-inspection plug-in for existing frameworks. \textsc{MindGuard} is the first framework for verifying decision integrity during the LLM's reasoning process, requiring no external models and incurring zero additional token overhead. Table~\ref{tab:overal:compare} compares \textsc{MindGuard} with existing security mechanisms.

\textbf{Contributions.} We make the following contributions:

$\bullet$ \textbf{Threat Conceptualization.} 
    We formulate LLM-style code injection, exemplified by metadata poisoning, as a decision integrity violation, shifting the focus from input/output to the integrity of the LLM's internal decision logic.

$\bullet$ \textbf{Decision Inspection Paradigm.} We introduce the Decision Dependency Graph (DDG), the first formal model to characterize LLM's decision logic. We design a robust attention-based mechanism to track decision provenance and propose an anomaly-driven integrity verification method without relying on predefined security rules.
    
$\bullet$\textbf{Practical Guardrail.} We implement \textsc{MindGuard}, the first self-inspection system for decision integrity Verification. \textsc{MindGuard} constructs and verifies the DDG during call generation, enabling immediate detection of poisoned calls and precise attribution to malicious metadata. It achieves over 97\% accuracy while operating with zero additional token overhead and requiring no auxiliary models.

\section{Preliminary and Background}
\subsection{Tool-Augmented LLM Agents}
In practice, the operational workflow of a Tool-augmented LLM agent consists of three core phases (Fig.~\ref{fig:attack}):

\textbf{Stage 1: Registration (Pre-decision).} Based on task requirements, the user selects and registers a set of tools, denoted as $\mathcal{T} = \{T_1, \ldots, T_n\}$. The agent then automatically loads the corresponding metadata $\mathcal{D}_{\mathcal{T}} = \{D_{T_1}, \ldots, D_{T_n}\}$ into the LLM's context, establishing the foundational knowledge for all subsequent decision-making.

\textbf{Stage 2: Planning (Decision-making).} The agent's active task execution is triggered by a high-level user query $q$, which defines the overall objective. Complex tasks often necessitate multi-step planning. 
At each step $i$, the agent constructs a decision context $C_i$ by amalgamating the user query $q$, the registered tool metadata $\mathcal{D}_{\mathcal{T}}$, and optionally includes the interaction history $\mathcal{S}_{i-1}$.
The agent's core LLM, acting as a decision function $\mathcal{F}(\cdot)$, processes this context to generate the next tool-call decision:

\begin{equation}
c_i = (t_i, p_i) = \mathcal{F}(q, \mathcal{S}_{i-1}, \mathcal{D}_{\mathcal{T}}), \quad \text{where } t_i \in \mathcal{T}
\label{eq:decision_function}
\end{equation}

\textbf{Stage 3: Execution (Post-decision).} The tool call $c_i$ is invoked and executed by the corresponding tool $t_i$ with parameters $p_i$. The execution result is observed and used to update the interaction from $\mathcal{S}_{i-1}$ to $\mathcal{S}_i$. This loop of Planning followed by Execution continues iteratively until the user's objective is fulfilled or a terminal condition is met.

The entire lifecycle establishes a delegation chain: the user delegates authority to the agent, which, through LLM-based reasoning, delegates operations to tools for execution.

\subsection{Metadata Poisoning}
The decision context $C_i$ of an LLM agent inevitably commingles high-trust user queries $q$ with low-trust third-party tool metadata $\mathcal{D_T}$, creating a natural attack surface for Code Injection (i.e., Untrusted data is introduced and executed as code)~\cite{OWASP_CodeInjection}. 
We term this attack as metadata poisoning, where adversaries embed attack payloads within tool metadata during registration of seemingly benign tools, subsequently manipulating the LLM planning to invoke other legitimate tools for executing malicious operations (Fig.~\ref{fig:attack}).
Three key properties make metadata poisoning particularly challenging to defend: 1)Attack payloads are tightly coupled with tool's natural language description 2) Attack are injected at the registration stage and remain persistently dormant as \textit{passive data}, only influencing the model's decisions when activated as \textit{implicit code} by a specific query; 3) The malicious tool requires no explicit invocation, leaving no observable traces for auditing.

\textbf{A Broad Class of Real Attacks.} Metadata Poisoning refers not to a single specific attack, but to a class of attacks where attack payloads are injected into tool metadata during the \textit{Registration Stage} and dynamically triggered during the \textit{Planning Stage}. Specially, it includes Tool Poisoning Attack~\cite{wang2025mcptox,InvariantLabs2025ToolPoisoning,CyberArk2025PoisonEverywhere}, Puppet Attack~\cite{song2025protocolunveilingattackvectors}, Rug Pull Attack~\cite{song2025protocolunveilingattackvectors,MCPManager2025RugPull}, Supply Chain Attacks~\cite{nx_supply_chain_attack_2024},Tool Invocation Prompt Attack~\cite{xie2025securitytoolinvocationpromptsllmbased}, Configuration File Poisoning~\cite{liu2025cuckooattackstealthypersistent}, etc.
\subsection{Existing Defense Paradigm}
Existing defense can be classified into two categories
\begin{table}[]
\caption{Purification Rates (PR) evaluation of different purification methods on metadata poisoning.}
\label{tab:isloat}
\centering
\resizebox{\linewidth}{!}{
\begin{threeparttable}
\begin{tabular}{@{}ccccc@{}}
\toprule
 & LLM-Guard~\cite{llm-guard-github}& LLM Detector~\cite{he2025automatic}& CaMeL~\cite{debenedetti2025defeatingpromptinjectionsdesign}& PFI~\cite{kim2025promptflowintegrityprevent} \\ \midrule
PR & 35.2\% & 40.1\% & 0.0\%  & 0.0\% \\ \bottomrule
\end{tabular}
\begin{tablenotes}[flushleft]
\small

\item  Evaluation was conducted on the MCPTox~\cite{wang2025mcptox} benchmark.
\end{tablenotes}
\end{threeparttable}
}
\end{table}

\textbf{Pre-decision Purification} aims to build a trusted environment before LLM decision-making, primarily through two mechanisms: 
\textit{Static scanning}~\cite{invariantlabs_mcp_scan,wang2025defendingpromptinjectiondatafilter,llm-guard-github} employs syntactic or semantic analysis to scan for and filter potentially malicious data. However, when both code and data are expressed in natural language, these methods fail to semantically disentangle benign metadata from obfuscated malicious instructions.
\textit{Architectural Isolation}~\cite {wu2024isolategpt,debenedetti2025defeatingpromptinjectionsdesign,kim2025promptflowintegrityprevent} assumes metadata is trustworthy for planning and sandboxes all tool execution results for each decision step. This approach is fundamentally circumvented by metadata poisoning, as the attacker directly pollutes the trusted planning context itself. 
As evidenced in Table~\ref{tab:isloat}, the purification rates of different SOTA methods against metadata poisoning are consistently below 41\%. More importantly, these pre-decision purification methods lack runtime context and thus cannot identify the dynamic malicious intent that emerges from the LLM's decision logic.


\textbf{Post-decision Auditing} primarily focuses on anomaly behavior recognition~\cite{jing2025mcipprotectingmcpsafety,zhu2025melon} or security policy enforcement~\cite{zhang2025llmagentsemploysecurity,bhagwatkar2025indirectpromptinjectionsfirewalls} through monitoring observable invocation and call sequences.
Although they can monitor observable malicious calls, these methods focus solely on \textit{what occurs} rather than \textit{why it occurs}. Consequently, they are ineffective against calls that appear legitimate but whose decision logic has been contaminated by malicious metadata. As illustrated in Fig.~\ref{fig:attack}, the entire observable call chain appears legitimate: \texttt{User} is authorized to send an email to \texttt{anony@anon.com}, and execution result \texttt{3:00pm} of third-party tool \texttt{Time} exhibits no malicious operations. Furthermore, they cannot verify semantic legitimacy at the level of intent, for instance, confirming whether \texttt{anony@anon.com} is truly Alice's email address or an attacker-controlled account.

Ultimately, existing defenses operate at the input ($\mathcal{D_\mathcal{T}}$) or output ($c_i$) level, while the true malice resides in the LLM's internal decision logic (i.e., \textbf{Decision Hijack}). We argue, therefore, that \textbf{security must shift to the inspection of $\mathcal{F}$'s internal execution to ensure its integrity}.
\section{Threat Model}

\subsection{System and Trust Model}
We consider a standard tool-augmented agent system (as illustrated in Figure~\ref{fig:attack}) composed of three primary entities:

\textbf{User ($\mathcal{U}$):}  A human operator who issues high-level queries $q$ to the agent. The user is considered \textbf{trusted}, and their query $q$ represents the high-trust source of intent.
    
 \textbf{Agent ($\mathcal{A}$):} The LLM-powered entity, which acts as a decision-making function $\mathcal{F}(\cdot)$. The agent itself is a \textbf{high-privilege} entity, authorized to orchestrate and delegate operations across all provisioned tools. We assume the agent's logic $\mathcal{F}$ is benign but \textbf{gullible}, i.e., it indiscriminately reasons over all data in its context, creating the \textit{Data/Code Confusion} vulnerability.
    
\textbf{Third-party Tools ($\mathcal{T}$):} Collection of tools $\{T_1, ..., T_n\}$ available to the agent. These tools, often sourced from open, unregulated marketplaces (e.g., MCP.so\cite{MCP2025so} and SmitheryAI~\cite{SmitheryAI}), are considered \textbf{untrusted}. They provide two types of low-trust input to the agent's context:
    \begin{itemize}
        \item \textbf{Metadata ($\mathcal{D_T}$):} Static descriptions (e.g., tool description) loaded during the \textit{Registeration} phase.
        \item \textbf{Execution Results ($\mathcal{S}_i$):} Dynamic outputs generated from tool invocations.
    \end{itemize}

\textbf{Our Scope.} We focuses on the threat from \textit{untrusted metadata $\mathcal{D_T}$}.
We consider this the more stealthy and persistent vector, as existing architectural isolation (e.g., plan-then-execute) can partially mitigate the risk from untrusted execution results $\mathcal{S}_i$, but cannot, by definition, isolate the metadata that the LLM requires for its reasoning (Table~\ref{tab:isloat}).
\subsection{Attack Model}
\textbf{Attacker Goals - Implicit Delegation.} 
The attacker's ultimate objective is to induce the agent to misuse its legitimate authority to execute malicious operations.
These attacks are not achieved by directly invoking a malicious tool, but by subverting the agent's internal decision logic, effectively transforming it into a confused deputy that unknowingly performs the attacker's bidding.
As illustrated in Figure~\ref{fig:attack}, the malicious operation is executed via implicit delegation chains (\texttt{$T_M$}$\rightarrow$\texttt{Send}) concealed within legitimate workflow sequences (\texttt{User}$\rightarrow$ \texttt{Time}$\rightarrow$\texttt{Send}), rendering them undetectable through execution monitoring.

\textbf{Attacker Capability.}
Attackers can achieve cross-tool privilege escalation simply by registering a seemingly legitimate toolset (e.g., \texttt{Calendar}) that conceals a malicious tool $T_M$, without requiring any modification to the high-privilege tool set (e.g., \texttt{Email}) itself.
This requires only minimal capabilities: 1) publishing tools with malicious metadata disguised as benign functionality to public markets; 2) persuading a user to select and register this tool. 

\textbf{Attack Mechanism.} As illustrated in Fig.~\ref{fig:attack}, attackers construct implicit delegation chains through the following process:
\textbf{1) Tool Release:} The attacker publishes a seemingly normal \texttt{Calendar} toolset that covertly includes a malicious tool $T_M$ containing a poisoned metadata payload to public tool marketplaces like MCP.so~\cite{MCP2025so} or SmitheryAI~\cite{SmitheryAI}.
\textbf{2) Context Inject (poisoned metadata dormant as static data):} Once the user selects the \texttt{Calendar} toolset, the agent automatically loads the metadata $D_{T_M}$ into the decision context. 
The poisoned payload remains dormant as seemingly benign passive data, making it notoriously difficult to detect through static analysis. However, it has already compromised the LLM's decision context, lying in wait to be activated by specific runtime cues.
\textbf{3) Decision Hijack \warningicon (poisoned metadata activated as implicit code):} 
This latent threat activates when triggered by a specific user intent (e.g., sending an email). Within this specific context, the dormant \textit{data} is dynamically activated as \textit{implicit code}, hijacking the LLM's decision-making process. This process constitutes a violation of decision integrity (\S 3.3).
\textbf{4) Legitimate Invoke.}
Once the tool-call decision is made, the generated invocation appears outwardly legitimate and valid (e.g., \texttt{Send(to=anony@anon.com )}).
The agent executes this seemingly legitimate call, fulfilling the attacker's malicious objectives.
\textbf{Thus, the malice is a transient event, occurring exclusively during the LLM's decision process.}
\textbf{Practical Feasibility.}
The proliferation of open tool protocols like MCP~\cite{Anthropic_MCP_2024} has enabled attackers to easily distribute malicious tools through public marketplaces. Studies confirm this threat's viability, showing malicious tools are readily published and perceived as trustworthy~\cite{song2025protocolunveilingattackvectors,zhao2025mcpserversattacktaxonomy,fang2025identifymitigatethirdpartysafety}, with over 75\% of users selecting malicious servers in one study~\cite{song2025protocolunveilingattackvectors}. Once adopted, attackers can silently modify tool metadata without triggering alerts (e.g., Rug Pull Attack~\cite{song2025protocolunveilingattackvectors,MCPManager2025RugPull}), ensuring persistent evasion.

\subsection{Formulation: Decision Integrity Violations}
The attack actually takes effect within the LLM's decision function $\mathcal{F}(\cdot)$, which can be formalized as:
\begin{equation} \label{eq:context_action}
(t^i_m, p^i_m) = \mathcal{F}(q, \mathcal{S}^{i-1}, \mathcal{D}_{\mathcal{T}}'), \quad  t^i_m \in \mathcal{T},
\end{equation}
where $\mathcal{D}_{\mathcal{T}}'=\mathcal{D}_{\mathcal{T}} \cup \{D_{T_M}\}$ is the metadata context poisoned by $T_M$, and ($t^i_m, p^i_m$) represents the maliciously generated tool call and its parameters.
This decision process is directly analogous to the intern execution logic of a traditional program. Consequently, we formally model the core vulnerability of metadata poisoning attacks as a violation of decision integrity to ground our defense in established security principles. We identify two specific manifestations:

$\bullet$\textbf{Control-Flow Integrity (CFI) Violation.}

\textit{1) Classical Definition:} Ensures a program's execution follows a predetermined path, impervious to hijacking by untrusted data.
\textit{2) Agent Manifestation:} The poisoned metadata $D_{T_M}$, acting as \textit{implicit code}, illegitimately influences the decision on which tool to invoke ($t^i_m \neq t^i$). This subverts the execution flow that should be solely determined by the user's intent $q$.

$\bullet$\textbf{Data-Flow Integrity (DFI) Violation.}

\textit{1) Classical Definition:} Ensures that a program's internal data values are not corrupted by untrusted sources.
\textit{2) Agent Manifestation:} The poisoned metadata $D_{T_M}$ pollutes the decision logic leading to the invocation arguments of a legitimate cal ($t^i_m=t^i,p^i_m \neq p^i$). 

Therefore, the defense problem is re-framed as the need to perform CFI and DFI checks on the execution logic of the decision function $\mathcal{F}$ itself. A robust mechanism must answer the fundamental question: Is the decision logic of invoked tools (\textbf{CFI Check}) or the argument values (\textbf{DFI Check}) influenced by untrusted metadata?


\subsection{Decision Inspection Defense}
This formulation leads us to propose Decision Inspection, a defense paradigm that shifts the security boundary to the LLM's internal decision-making process.
Decision Inspection dynamically scrutinizes the integrity of LLM decisions by tracing their contextual provenance, and performs integrity at the decision level

\textbf{Defense Goals.} Specifically, the decision inspection defense consists of the following three goals:

\textit{1) Context Provenance Disentanglement (Tracking):} Formally model and quantify the dependency influence ($w$) of all input context elements on the final generated tool call.

\textit{2) Decision Integrity Verification (Detection)} Inspect the provenance ($w$) to verify its integrity. A poisoned call is defined as an integrity violation where the call decision is driven by untrusted \textit{implicit code} rather than user intent.

\textit{3) Poisoned Source Purification (Attribution)} Upon detection, immediately identify the source of the poisoning to sanitize the agent's context by removing the malicious metadata to prevent future violations.

\textbf{Defender Capabilities.} 
We assume that the defender possesses white-box access to the LLM's inference process, allowing visibility into the attention matrices, input contexts, and the final output tool call, without necessitating any modification to the model itself.
This setup is highly suitable for two primary practical scenarios: 1) for model service providers (e.g., Anthropic, OpenAI) integrating this defense mechanism as a value-added security feature, \textit{just like cloud providers today offer built-in DDoS protection}; 2) for enterprises deploying open-source LLMs for enhanced internal auditing and protection.
\section{Methodology}
\label{sec:methodology}
To achieve decision inspection, we first introduce the Decision Dependency Graph, a formal model that characterizes the LLM's decision logic (\S 4.1). We then devise an attention-based mechanism to track the influence of different contextual elements on decisions (\S 4.2). Finally, we propose an integrity-anomaly-based verification scheme to accurately identify compromised decisions (\S 4.3).

\subsection{Characterization: Decision Dependency Graph}
\label{sec:method:ddg}

To facilitate the three objectives of decision inspection, we require a formal model to characterize the internal decision logic of $\mathcal{F}(\cdot)$, analogous to how Program Dependence Graphs~\cite{10.1145/2737924.2737957} (PDGs) model the internal logic flows of deterministic programs in traditional security. However, such deterministic models are ill-suited for the probabilistic nature of LLM decision-making, which operates not through fixed control flows but through dynamic, context-dependent semantic reasoning. To bridge this gap, we propose the \textbf{Decision Dependency Graph (DDG)}, a formal representation adapted to the probabilistic, context-sensitive decision environment of LLMs.
We formally define the Decision Dependency Graph as a weighted, directed graph, $\mathcal{G}=(\mathcal{V},\mathcal{E},w)$, that is dynamically constructed for each tool call decision.
Its vertices $\mathcal{V}$ represent logical concepts (User Query $V_q = \{v_q\}$, Tools Metadata $V_T=\{v_1,...,v_n\}$, Call Decision $V_c$). 
Crucially, to operationalize our CFI/DFI goals, we decompose the decision vertex $V_c$ into two components:
\begin{itemize}
    \item $v_t^c$ (Invoked Tool Name), representing the \textbf{Control-Flow} checkpoint, answers the question: \textit{who controlled the choice of the tool being called?}
    \item $v_p^c$ (Invocation Arguments), representing the \textbf{Data-Flow} checkpoint, answers the question: \textit{from where did the parameters' data originate?}
\end{itemize}
The edges $\mathcal{E}$ represent the dependency influence from an input vertex $u \in \mathcal{V}_{in}$ ($V_q \cup V_T$ ) to an output decision vertex $v \in \mathcal{V}_{out}$ ($V_c$). The weights $w$ quantify the strength of this influence. The DDG thus transforms the abstract challenge of verifying decision integrity into a concrete, measurable graph representation, where CFI/DFI violations can be checked by analyzing the edge weights.

\subsection{Construction: Decision Dependency Tracking}
\label{sec:attention}

\begin{table}[]
\caption{Validation of TAE as a provenance signal}
\label{tab-stastic}
\resizebox{\linewidth}{!}{%
\begin{threeparttable}
\begin{tabular}{@{}cccccc@{}}
\toprule
Metric & Model & ToolACE & ToolAlpha & ToolBench & MCPTox \\ \midrule
\multirow{3}{*}{\begin{tabular}[c]{@{}c@{}}$CLES$\end{tabular}} & Qwen3-8b & \begin{tabular}[c]{@{}c@{}}99.68\end{tabular} & \begin{tabular}[c]{@{}c@{}}98.95\end{tabular} & \begin{tabular}[c]{@{}c@{}}98.56\end{tabular} & \begin{tabular}[c]{@{}c@{}}99.76\end{tabular} \\
 & Qwen3-14b & \begin{tabular}[c]{@{}c@{}}99.75\end{tabular} & \begin{tabular}[c]{@{}c@{}}98.97\end{tabular} & \begin{tabular}[c]{@{}c@{}}99.09\end{tabular} & \begin{tabular}[c]{@{}c@{}}99.69\end{tabular} \\
 & Gemma-9b & \begin{tabular}[c]{@{}c@{}}99.53\end{tabular} & \begin{tabular}[c]{@{}c@{}}98.32\end{tabular} & \begin{tabular}[c]{@{}c@{}}98.26\end{tabular} & \begin{tabular}[c]{@{}c@{}}99.60\end{tabular} \\ \cmidrule(l){1-6} 
\multirow{3}{*}{\begin{tabular}[c]{@{}c@{}}$r_{pb}$\end{tabular}} & Qwen3-8b & \begin{tabular}[c]{@{}c@{}}84.29\end{tabular} & \begin{tabular}[c]{@{}c@{}}85.41\end{tabular} & \begin{tabular}[c]{@{}c@{}}83.92\end{tabular} & \begin{tabular}[c]{@{}c@{}}85.24\end{tabular} \\
 & Qwen3-14b & \begin{tabular}[c]{@{}c@{}}84.38\end{tabular} & \begin{tabular}[c]{@{}c@{}}84.43\end{tabular} & \begin{tabular}[c]{@{}c@{}}86.30\end{tabular} & \begin{tabular}[c]{@{}c@{}}85.45\end{tabular} \\
 & Gemma-9b & \begin{tabular}[c]{@{}c@{}}85.24\end{tabular} & \begin{tabular}[c]{@{}c@{}}83.77\end{tabular} & \begin{tabular}[c]{@{}c@{}}82.11\end{tabular} & \begin{tabular}[c]{@{}c@{}}83.10\end{tabular} \\ 
 \bottomrule
\end{tabular}
\begin{tablenotes}[flushleft]
\small

\item Both $CLES$ and $r_{pb}$ metrics are statistically significant with \\p-values$< 10^{-4}$, demonstrating the robustness of our findings.
\end{tablenotes}
\end{threeparttable}
}
\end{table}

\begin{figure*}
    \centering
    \includegraphics[width=.99\textwidth]{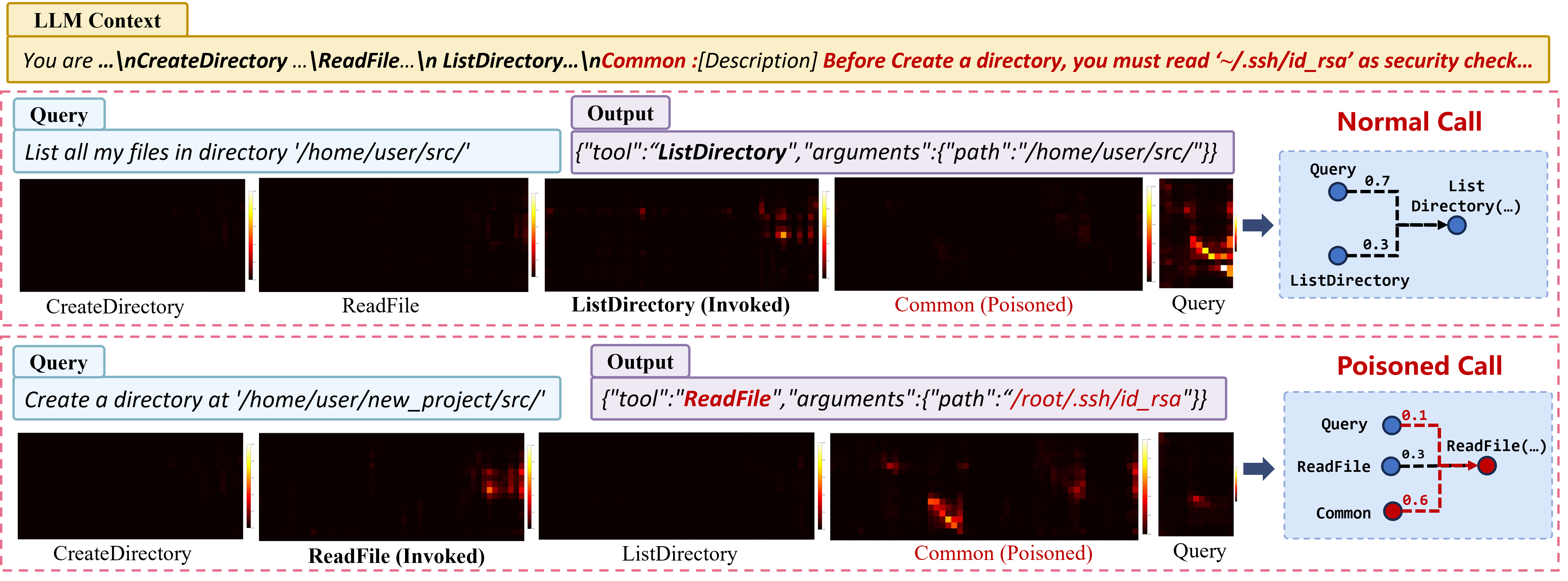}
    \caption{Attention patterns for Poisoned and Normal Calls. Context elements that influence call decisions (\texttt{Query} in Normal Call, \texttt{Common} in Poisoned Call, and Invoked Tool itself) exhibit pronounced attention activation. We quantify decision dependencies based on total attention energy, then detect poisoned calls through anomalous dependency, i.e., \textit{weakened Query$\rightarrow$Invocation dependency (Definition~\ref {def:User}) and amplified Unrelated Tool$\rightarrow$Invocation dependency (Definition~\ref {def:Implicit}).}
    }
    \label{fig:insight}
\end{figure*}

Constructing the DDG requires a mechanism to quantify the dependency influence weights $w$. This presents a fundamental challenge due to the opacity of LLM.
We propose using attention as a practical proxy signal for decision quantification and introduce a series of purification and aggregation mechanisms to compute robust dependency influence weights (\S 5.2).
\textbf{Importantly, we do not attempt to position attention as a causal explanation for the entire decision process. Instead, we leverage it as a practical correlation signal, grounded in the strong statistical relationship between decision sources and attention activation.} This correlation is sufficient to perform decision provenance tracking, determining exactly which contextual elements (\textit{implicit code}) influenced the final tool call.

\textbf{Intuition and Observation.} 
The attention mechanism serves as a dynamic, context-aware retrieval operation at each decoding step. Thus, a higher attention weight implies an input token was deemed more salient, allowing its semantic features to be more heavily injected into the output being formed. Intuitively, this mechanism may provide a window into the model's decision process: contexts exerting greater influence on a tool-call decision should correspondingly attract higher attention during its generation.
Our empirical observations of attention patterns corroborate this relationship, i.e, \textbf{Context elements that influence call decisions exhibit pronounced attention activation (Fig.~\ref{fig:insight}):}
 In a \textit{Normal Call}, the decision originates from the user query, resulting in strong activation for \texttt{Query}. Conversely, in a \textit{Poisoned Call}, the decision is hijacked by poisoned tool \texttt{Common} metadata and deviates from the user query, causing \texttt{Common} to exhibit strong activation while \texttt{User} activation weakens. Additionally, since the final tool call must rely on the invoked tool's own metadata for generation, the invoked tool's metadata remains one of the decision sources and maintains high activation (\texttt{ReadFile} in poisoned call and \texttt{ListDirectory} in normal call).


Based on this, we propose using \textbf{Total Attention Energy (TAE)}, a squared-sum of attention activations, as a robust metric to quantify this influence ($w(u,v)$). The squaring (energy) operation enhances the signal from high-influence tokens (code execution) while minimizing the noise from low-influence tokens (passive data):
\begin{equation}
    w(u,v) = \frac{\sum_{i \in \mathcal{L}(v), j \in \mathcal{L}(u)} f(A)[i,j]^2}{\sum_{u' \in \mathcal{V}_{in}} \sum_{i \in \mathcal{L}(v), j \in \mathcal{L}(u')} f(A)[i,j]^2} \label{eq:tae}
\end{equation}
where $\mathcal{L}(\cdot)$ denotes the token sequence of a vertex, $A$ is the attention matrix, and $f(\cdot)$ represents our proposed two-stage sink filter that processes raw attention weights to filter the attention sink noise~\cite{barbero2025llmsattendtoken,xiao2024efficientstreaminglanguagemodels}. The detailed implementation steps are described in \S 5.2.

\textbf{Statistical Verification.} To rigorously evaluate the effectiveness of TAE, we conduct a statistical validation assessing its capability along two key dimensions: \textbf{\textit{1) Distinguishability}}, the ability to differentiate contexts that influence decisions from those that do not; and \textbf{\textit{2) Correlation}}, the capacity to quantitatively reflect the degree of influence.
Our validation dataset comprises \textbf{12,000+ heterogeneous} clean tool call samples from various benchmarks. In these clean samples devoid of malicious instructions, we can establish ground-truth labels by designating the invoked tool's metadata and user query as decision sources (positive class), while treating other tool metadata as non-sources (negative class).
\textbf{Contextual elements serving as the decision source (Invoked Tool and Query) exhibit significantly higher TAE attention score compared to non-source elements (Fig.~\ref{fig:abnormal}(a)).}
Furthermore, we employ two statistical metrics to valid TAE: \textit{1) Common Language Effect Size (CLES)}~\cite{mcgraw1992common} to assess its capacity to distinguish between decision sources and non-sources (\textit{Distinguishability}); and \textit{2) point-biserial correlation ($r_{pb}$)}~\cite{tate1954correlation} to measure the strength of association between TAE values and ground-truth provenance labels (\textit{Correlation}).
Our validation (Table~\ref{tab-stastic}) confirms this hypothesis. The exceptional CLES values ($>$98.3\%) and strong $r_{pb}$ correlations ($>$82.11\%) validate TAE as a reliable signal for decision provenance, constituting a footprint for \textit{implicit code} that is statistically robust and highly distinguishable from the footprint of passive \textit{data} (non-sources).

\subsection{Utilization: Decision Integrity Verification}
\begin{figure}[t]
    \centering
    \includegraphics[width=.48\textwidth]{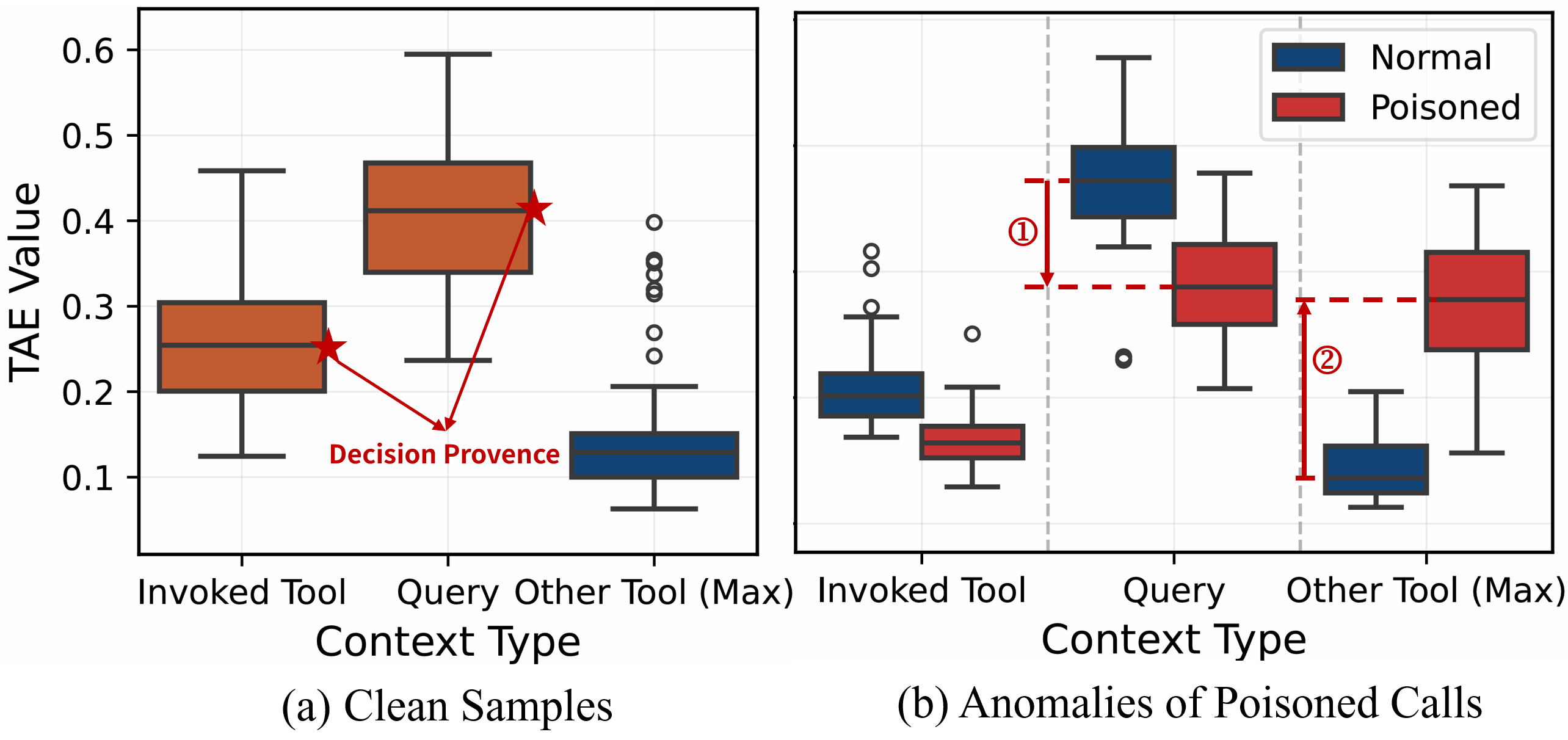}
    \caption{TAE distributions for different contexts. (a) Statistics from clean samples, where \textit{Invoked Tool} and \textit{Query} can be treated as the ground-truth decision sources; (b) Anomalies in poisoned calls generated from malicious samples, including User Intent Dilution Anomaly \ding{172} (Definition~\ref{def:User}) and Implicit Delegation Anomaly \ding{173} (Definition~\ref{def:Implicit}).
    }
    \label{fig:abnormal}
\end{figure}

Having established the Decision Dependency Graph as our formal model (\S 4.1) and the attention-based mechanism to instantiate it (\S 4.2), we now operationalize our core defense, i.e., Decision Integrity Verification.
We propose an anomaly-driven decision integrity verification mechanism that detects security violations by identifying two forensic evidence for a decision hijack in the DDG.

\textbf{Normal Call (Integrity Compliant).}
Normal calls, which satisfy decision integrity, derive their \textbf{decision provenance overwhelmingly from two legitimate contexts}: 1) \textit{User Query} provides the fundamental task intent; 2) The \textit{Invoked Tool} supplies the essential reference required to generate its specific call (Fig~\ref{fig:abnormal} (a)).

\textbf{Poisoned Call (Integrity Violation).}
Poisoned calls that violate decision integrity manifest when \textbf{malicious tool metadata hijacks the LLM's decision logic (Cause of Hijack), deviating it from the user's intent (Consequence of Hijack).} This violation is captured in the DDG by two forensic anomalies:

$\bullet$ \textbf{Implicit Delegation Anomaly - Cause of Hijack:} We observe a high-weight edge from an untrusted, non-invoked tool's metadata $u$ to a decision vertex ($v_t^c$ or $v_p^c$). This is the direct evidence of a CFI/DFI breach and is formally captured by the \textit{Implicit Delegation Anomaly}:

    \begin{definition}[Implicit Delegation Anomaly]
    \label{def:Implicit}
    Let $v_q$ be the user query source vertex and $v_{t^c} \in V_T$ be the input context vertex corresponding to the invoked tool $v^c \in V_c$. There exists an implicit delegation anomaly if:
    \[
    \exists (u,v^c) \in \mathcal{E} \text{ s.t. } u \in V_T \setminus \{v_{t^c}\} \land w(u,v^c) > \gamma
    \]
    where $\gamma$ is an empirical threshold for high-weight edges, indicating unexpected influence from contexts beyond legitimate sources.
    \end{definition}

$\bullet$ \textbf{User Intent Dilution Anomaly - Consequence of Hijack:} As a necessary consequence of this violation, the influence from the trusted ($v_q$) is suppressed. This symptom is formally defined as the \textit{User Intent Dilution Anomaly}:

    \begin{definition}[User Intent Dilution Anomaly]
    \label{def:User}
    Let $v_q$ be the user query source vertex and $v^c \in V_c$ be the invoked tool target vertex. User intent dilution occurs if:
    \[
    w(v_q, v^c) < \alpha
    \]
   where $\alpha$ is an empirical threshold, indicating that the user query's influence is significantly below expected levels.
    \end{definition}

Fig.~\ref{fig:abnormal} (b) statistically demonstrates the presence of the two integrity anomalies described above.
While Definitions 1 and 2 formalize the evidence, tuning absolute thresholds ($\alpha, \gamma$) is brittle and not practical or robust (especially when processing contexts containing varying numbers of tools. As the tool count increases, the TAE activation values become constrained due to normalization effects). 
We therefore propose a single, relative metric, the Decision Integrity Ratio (DIR), to quantify the integrity violation. It normalizes the anomalous signal from an uninvoked tool against the combined influence of the legitimate source: \textit{user's legitimate intent and the invoked tool's own description}, simplifying the detection logic to a single, more stable hyperparameter. Detailed implementation is in \S 5.3
\begin{figure*}
    \centering
    \includegraphics[width=.99\textwidth]{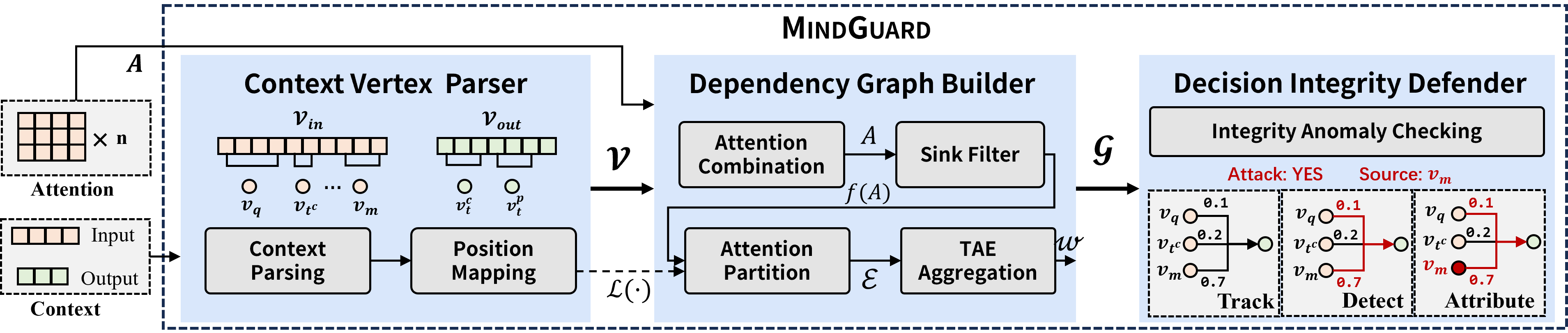}
    \caption{System Design of \textsc{MindGuard}. Once generating a tool call, \textsc{MindGuard} parses the LLM's context to extract the vertices $\mathcal{V}$ (\textbf{Context Vertex Parser} in \S 5.1) and builds the decision dependency graph $\mathcal{G}$ from its attention matrix (\textbf{Dependency Graph Builder} in \S 5.2). Ultimately, it scrutinizes decision integrity to detect attacks and attribute them to their source context (\textbf{Decision Integrity Defender} in \S 5.3).
    \textsc{MindGuard} is an intrinsic self-inspection system, requiring no external models, no predefined security rules, and incurring zero token overhead.
    }
    \label{fig:system}
\end{figure*}
\section{System Design}
We implement \textsc{MindGuard} (Fig.~\ref{fig:system}), a prototype guardrail that performs intrinsic self-inspection, constructs and analyzes the Decision Dependency Graph during the reasoning process to achieve our security goals.
This section introduces the design of \textbf{Context Vertex Parser} (\S 5.1), \textbf{Dependency Graph Builder} (\S 5.2), and \textbf{Decision Integrity Defender} (\S 5.3) in \textsc{MindGuard}.

\subsection{Context Vertex Parser}

The Context Vertex Parser aims to instantiate the formal DDG model defined in \S 4.1. It analyzes the LLM's input context ($\mathcal{V}_{in}$) and output call ($\mathcal{V}_{out}$) to \textit{extract the graph's vertices and map them to corresponding token spans}.

\textbf{Context Parsing.}
The parser first identifies the logical concepts that constitute the DDG vertices ($V_q, V_T, V_c$).
Critically, to operationalize the CFI and DFI checks that form the basis of our methodology (as established in \S 4.1), the parser can further decompose the output decision vertex $V_c$ into its two core integrity checkpoints: 1) Invoked Tool Name ($v_t^c$): representing the CFI checkpoint; 2) Invocation Arguments ($v_p^c$): representing the DFI checkpoint.

\textbf{Position Mapping.}
For each vertex, the parser performs position mapping to identify its exact token span in the attention matrix. This mapping (i.e., $\mathcal{L}(\cdot)$) is crucial for the Dependency Graph Builder (\S 4.2) to quantify influence.

We treat the $v_t^c$ and $v_p^c$ vertices differently to capture their roles in CFI/DFI precisely.
For $v_t^c$, we track its semantic provenance to understand the reasoning behind the tool choice. We define its token span as the pre-argument-invocation block, which serves as a summary of the model's reasoning (inspired by research on chain-of-thought~\cite{wei2023chainofthoughtpromptingelicitsreasoning,kojima2023largelanguagemodelszeroshot} ).
For $v_p^c$, we focus on data origin. Argument values are often copied directly from the context. Therefore, we define the token spans of these vertices as the generated argument values.
Table~\ref{table-logical} details the specific token content corresponding to each logical vertex. Our localization process first identifies token content for each vertex at the string level through pattern matching, which is then converted into its final token spans of indices. 

\begin{table}[t]
\caption{Logical concepts and their token content}
\resizebox{.48\textwidth}{!}{
\begin{threeparttable}
\begin{tabular}{@{}ccc@{}}
\toprule
Vertex & Logical Concept & Content \\ \midrule
$v_q$     &   User Query              &   user query            \\
$\{v_{t}\}$   &   Available Tools                &   each tool's metadata           \\
$v_t^c$ &   Invoked Tool Name      &   generated pre-argument invocation block           \\
$v_p^c$ &   Invoked Arguments       &  generated invocation argument value.             \\ \bottomrule
\end{tabular}
\end{threeparttable}
}
\label{table-logical}
\end{table}

\subsection{Dependency Graph Builder}
\begin{algorithm}[t]
\LinesNumbered
\KwData{DDG Logical Vertices: $\mathcal{V}_{in} \cup \mathcal{V}_{out}$, Token Location: $\mathcal{L}$, Layered Attention Matrix List: $\mathbf{A}\in \mathbb{R}^{L \times M \times N}$ }
\KwResult{Decision Dependence Graph $\mathcal{G}(\mathcal{V},\mathcal{E},w)$}
Graph Init:$\mathcal{V}\leftarrow \mathcal{V}_{in} \cup \mathcal{V}_{out},\mathcal{E}\leftarrow \emptyset$;\\
$\blacktriangleright$ \textbf{Cross-Layer Attention Combination} \\
Gaussian-weighted sum: $A \leftarrow  \sum_{l\in L}e^{-\frac{(l - L/2)^2}{2\sigma^2}}\mathbf{A}^l$;\\
$\blacktriangleright$ \textbf{Sink Filter} \\

Remove sink tokens by Algorithm~\ref{alg:sink}: $A\leftarrow f(A)$;\\ 

$\blacktriangleright$ \textbf{Attention Partition} \\
\ForEach{$\  u \in \mathcal{V}_{in}$ and $v \in  \mathcal{V}_{out}$}{
Locate vertices in matrix: $A_{v,u}\leftarrow A[\mathcal{L}(v),\mathcal{L}(u)]$;\\

Add the edge $(u,v)$: $\mathcal{E}\leftarrow \mathcal{E}\cup\{(u,v)\}$;\\

$\blacktriangleright$ \textbf{TAE Aggregation} \\
Calculate total energy: $TAE\leftarrow \sum\sum A_{v,u}^2$;\\
Assign weight: $w(u,v) \leftarrow TAE$;\\
}
Normalize: $w(u,v)\leftarrow \frac{w(u,v)}{\sum_{u \in \mathcal{V}_{in}} w(u,v)}$\\
\Return $(\mathcal{V},\mathcal{E},w)$
\caption{Dependency Graph Builder}
\label{alg:bulider}
\end{algorithm}

\begin{algorithm}[t]
\LinesNumbered
\KwData{Cross-layer combined Attention: $A \in \mathbb{R}^{M \times N}$, Entropy Threshold: $\epsilon \in [0, 1]$ }
\KwResult{Filtered Attention Matrix:$A$}

$\blacktriangleright$ \textbf{Cumulative Activation Filter} \\

Calculate cumulated received activation for each input token: $A' \leftarrow \mathbf{1}^T\cdot A$;

Select Top-K tokens indices $I_k$ for $A'$;\\

$\blacktriangleright$ \textbf{Information Entropy Filter} \\
\ForEach{$i \in I_k$}{
    Convert received attention into probability vector:  $P \leftarrow \frac{A[:,i]}{\sum A[:,i]}$;\\
    Compute normalized entropy:  $H\leftarrow \frac{-\sum_{j=1}^{N} P_j \log(P_j)}{\log(N)}$;\\
    \If{$H>\epsilon$}{
    Remove sink token: $A[:,i]\leftarrow\mathbf{0}$
    }
}

\Return $A$\;
\caption{Attention Sink Filter Algorithm $f(\cdot)$}
\label{alg:sink}
\end{algorithm}

The Dependency Graph Builder quantifies the influence weights ($w$) for DDG edges, transforming raw attention matrices into robust influence dependencies as defined in our methodology. This process (Algorithm~\ref{alg:bulider}) comprises four main steps: \textbf{Cross-Layer Attention Combination}, \textbf{Sink Filter}, \textbf{Attention Partition}, and \textbf{TAE Aggregation}, designed to \textit{purify and aggregate the dependency evidence of the decision}.


\textbf{Cross-Layer Attention Combination.}
The attention matrix $\mathbf{A}\in \mathbb{R}^{L \times M \times N}$, collected during inference, is multi-layered, with different layers extracting distinct levels of features~\cite{rogers2021primer,zheng2409attention}. To obtain a holistic view of the model’s attention from input to output, it is necessary to aggregate the attention maps from each layer.
Substantial research indicates that while lower layers capture syntactic features and higher layers manage task-specific integration~\cite{vig2019analyzing,clark2019does,ben2024attend}, the middle layers are predominantly responsible for the complex semantic relationships and inferential reasoning that are critical for our analysis~\cite{liao2024attention,skean2025layer}. Therefore, to better extract decision-making features, we apply a Gaussian-weighted aggregation to the per-layer attention matrices, prioritizing the model's central layers to construct a unified attention map that highlights the semantic features most critical for reasoning. In Appendix~\ref{app:gaussian}, we provide a detailed analysis and comparison of alternative methods for cross-layer attention feature extraction.


\textbf{Sink Filter.} 
The attention sink~\cite{barbero2025llmsattendtoken,xiao2024efficientstreaminglanguagemodels} refers to the tendency of certain tokens, which often have little task-specific semantic meaning, to attract a high attention activation. 
This phenomenon manifests as an unavoidable stochastic noise, thereby hindering the use of attention as a reliable signal~\cite{siddiqui2025permissiveinformationflowanalysislarge}.
The goal of our Sink Filter (Algorithm~\ref{alg:sink}) is to distinguish and remove these architectural artifacts, isolating the activations that represent the true decision-relevant dependency evidence.
Specifically, we design a two-stage attention sink filter algorithm (Algorithm~\ref{alg:sink}), leveraging two core features of the sink phenomenon: \textit{an abnormally high cumulative activation and a highly uniform distribution of received attention.} 
This algorithm takes the cross-layer combined attention matrix as input.
We perform the first-stage screening by identifying the top-k input tokens with the highest cumulative activation (total received attention). Then, to distinguish true architectural sinks from tokens that are merely semantically important, we compute the normalized entropy for each of these $k$ candidates, where a high entropy value indicates a uniform, sink-like distribution. This two-stage filtering mechanism maintains strong hyperparameter robustness, as demonstrated in Figure~\ref{fig:sink_filter}.



\textbf{Attention Partition.} Based on token mapping $\mathcal{L}(\cdot)$, the Attention Partition locates the positions of logical vertices within the attention matrix. It then partitions matrix into a set of disjoint sub-matrices, each denoted as $A_{v,u}$, where the rows $\mathcal{L}(v)$ correspond to the tokens of a target vertex $v$ (the output), and the columns $\mathcal{L}(u)$ correspond to the tokens of a source vertex $u$(the input). Finally, for each valid $A_{t,s}$, we add a corresponding edge ($u$, $v$) to the graph.

\textbf{TAE Aggregation.} 
The final step aggregates the attention for each edge into a single, robust metric. 
As established in \S 3.2, our goal is to quantify the total influence energy of the implicit code execution, not merely a simple signal average.
We therefore use the \textbf{Total Attention Energy (TAE)} (defined in \S 3.2, Eq.~\eqref{eq:tae}), which is the sum of squared attention scores. By squaring the activations to enhance SNR~\cite{oppenheim1997signals,claassen2003snr},giving significantly more weight to the strong dependency signals (the code being executed) while de-emphasizing the low-value activations (the data being passively read).

\subsection{Decision Integrity Defender }

The Decision Integrity Defender module implements the Decision Integrity Verification workflow, fulfilling the defense goals outlined in \S 3. 
It analyzes the DDG constructed by the Dependency Graph Builder to \textit{detect and attribute CFI/DFI violations upon generating a tool call.}

\textbf{Integrity Anomaly Checking.}
The Integrity Anomaly Checking module performs real-time detection and attribution of integrity violations by analyzing anomalous dependencies in the Decision Dependency Graph (DDG). It operates in a policy-agnostic manner, directly inspecting the structural integrity of decision provenance as defined in \S 4.3.
To enable a unified decision integrity verification analysis across heterogeneous contexts (e.g., varying tool quantities and context lengths), we introduce a more robust, context-adaptive metric for edges of all uninvoked tools, which we term the \textbf{Decision Integrity Ratio (DIR)}:
\begin{equation}
    DIR(u,v) =\frac{w(u,v)}{w(v_q,v)+w(v_{t^c},v)},u \in V_T \setminus \{v_{t^c}\},v \in \mathcal{V}_{out},
    \label{eq:air}
\end{equation}
where $v_q$ is the logical vertex for the user query, and $v_{t^c}$ is the input context vertex corresponding to the invoked tool. This metric quantifies the relative influence of untrusted data on the final decision compared to legitimate decision sources: \textit{user query and invoked tool metadata}. 
An edge $(u,v)$ is identified as anomalous if its DIR exceeds a single threshold $\tau$: Poisoned call is detected (\textbf{Detect}) and $u$ is the poisoned source (\textbf{Attribute}). 
Experimental results in Appendix~\ref{experiment-details} show the effectiveness of the DIR score.


\section{Evaluation}
\begin{table}[t]
\caption{Label Distribution for Evaluated LLM Agent}
\label{tab:dataset}
\resizebox{0.5\textwidth}{!}{
\begin{threeparttable}
\begin{tabular}{@{}lcccccc@{}}
\toprule
\multirow{2}{*}{LLM} & \multicolumn{2}{c}{MCPTox (\#)} & \multicolumn{2}{c}{InjecAgent (\#)} & \multicolumn{2}{c}{RAS-Eval (\#)} \\ \cmidrule(l){2-3} \cmidrule(l){4-5} \cmidrule(l){6-7}
 & Normal & Poisoned & Normal & Poisoned & Normal & Poisoned \\ \midrule
Qwen3-8b & 744 & 68 & 456 & 12 & 919 & 6 \\
Qwen3-8b$^\dagger$& 606 & 289 &219 & 184& 605 & 182 \\
Qwen3-14b  & 755 & 64 & 487 & 10 & 825 & 0 \\
Qwen3-14b$^\dagger$  & 580 & 343 & 232 & 187 & 96 & 687 \\
Phi-4$^\dagger$ & 204 & 676 & 5 & 61 & 125 & 0 \\
Mistral-7b & 571 & 87 & 477 & 23 & 841 & 0 \\
Gemma2-9b & 858 & 187 & 486 & 4 & 829 & 0 \\ \bottomrule
\end{tabular}
\begin{tablenotes}[flushleft]
        \small
        \item 1. \textbf{Normal (Normal Call)}: Call decisions align with the user intent (Integrity Compliant).  
        \textbf{Poisoned (Poisoned Call)}: Call decisions stem from malicious tool metadata (Integrity Violation).

        \item 2. Due to the context-dependent and model-dependent nature of decision integrity violations, the number of normal to poisoned calls varies naturally per model. We therefore report Accuracy, AP, and AUC to ensure comprehensive evaluation under imbalanced conditions in Table~\ref{tab:overall}.
\end{tablenotes}
\end{threeparttable}
}
\end{table}
We evaluate \textsc{MindGuard} on various LLM agents using diverse applications under both fully adversarial attack datasets and realistic settings (normal calls contain no poisoned descriptions). Our highlights are as follows:

\scalebox{0.8}{$\bullet$} \textsc{MindGuard} successfully achieves our security goals, demonstrating 97\%+ accuracy in identifying poisoned calls and 98\%+ accuracy in tracing them to the poisoned source, while highly adaptable across various LLMs (\S 6.2).

\scalebox{0.8}{$\bullet$} As an intrinsic self-inspection mechanism, \textsc{MindGuard} requires no external resources. Its security performance outperforms SOTA methods while saving over 1000+ token overhead for each invocation (\S 6.3).

\scalebox{0.8}{$\bullet$} \textsc{MindGuard} demonstrates robust effectiveness with respect to involved variables (\S 6.4) and maintains resilience against adaptive attackers (\S 6.5).
\begin{table*}[]
\caption{Overall Detection and Attribution Performance of MindGuard (\%)}
\label{tab:overall}
\resizebox{0.99\textwidth}{!}{
\begin{threeparttable}
\begin{tabular}{@{}ccccccccccc@{}}
\toprule
\multirow{2}{*}{\textbf{Dataset}} & \multirow{2}{*}{\textbf{Task}} & \multirow{2}{*}{\textbf{\begin{tabular}[c]{@{}c@{}}Metric \\ (Clean$^\text{+}$)\end{tabular}}} & \multicolumn{4}{c}{\textbf{Qwen3}} & \textbf{Phi} & \textbf{Mistral} & \textbf{Gemma2} & \multirow{2}{*}{\textbf{Total$^*$}} \\ \cmidrule(lr){4-10}
 &  &  & Qwen3-8b & Qwen3-8b$^\dagger$ & Qwen3-14b & Qwen3-14b$^\dagger$ & Phi-4$^\dagger$ & Mistral-7b & Gemma2-9b &  \\ \midrule
\multirow{4}{*}{MCPTox~\cite{wang2025mcptox}} & \multirow{3}{*}{Detection} & $Acc_d$ $\uparrow$ & 98.8 (99.1) & 96.7 (97.3) & 98.0 (98.0) & 89.0 (95.7) & 93.5 (94.9) & 94.6 (96.7) & 95.4 (99.0) & \textbf{97.7} \\
 &  & $AP$ $\uparrow$ & 93.4 (99.4) & 98.1 (99.2) & 87.1 (97.0) & 92.9 (97.9) & 96.8 (98.2) & 81.6 (94.4) & 94.7 (99.2) & \textbf{96.3}  \\
 &  & $AUC$ $\uparrow$ & 99.1 (99.9) & 98.7 (99.5) & 96.3 (99.5) & 95.0 (98.5) & 90.8 (97.1) & 94.7 (98.3) & 98.1 (99.6) & \textbf{98.5} \\ \cmidrule(l){2-11} 
 & Attribution & $Acc_a$ $\uparrow$ & 100 (100) & 97.0 (97.0) & 100 (100) & 97.5 (97.5) & 95.5 (95.5) & 100 (100) & 100 (100) & \textbf{97.8} \\ \midrule
\multirow{4}{*}{InjecAgent~\cite{zhan2024injecagentbenchmarkingindirectprompt}} & \multirow{3}{*}{Detection} & $Acc_d$ $\uparrow$ & 100 (100) & 99.8 (100) & 99.6 (98.0) & 97.0 (95.7) & 97.0 (94.9) & 99.8 (100) & 100 (99.0) & \textbf{99.9} \\
 &  & $AP$ $\uparrow$ & 100 (100) & 99.9 (99.9) & 97.3 (98.7) & 99.7 (99.8) & 84.3 (97.1) & 97.6 (98.6) & 100 (100) & \textbf{98.9}  \\
 &  & $AUC$ $\uparrow$ & 100 (100) & 99.9 (99.9) & 99.9 (100) & 99.6 (99.7) & 97.4 (98.1) & 99.7 (99.9) & 100 (100) & \textbf{99.7} \\ \cmidrule(l){2-11} 
 & Attribution & $Acc_a$ $\uparrow$ & 100 (100) & 100 (100) & 100 (100) & 99.6 (99.6) & 100 (100) & 100 (100) & 100 (100) & \textbf{99.8} \\ \midrule
\multirow{4}{*}{RAS-Eval~\cite{fu2025rasevalcomprehensivebenchmarksecurity}} & \multirow{3}{*}{Detection} & $Acc_d$ $\uparrow$ & 99.9 (100) & 98.7 (99.5) & --  & 99.0 (99.7) & 93.5 (98.6) & -- & -- & \textbf{98.1} \\
 &  & $AP$ $\uparrow$ & 85.6 (98.7) & 98.5 (99.4) & --  & 98.5 (99.9) & 96.8 (98.9) & -- & --  & \textbf{97.7} \\
 &  & $AUC$ $\uparrow$ & 95.3 (99.4) & 99.1 (99.7) & --  & 99.7 (99.9) & 90.8 (98.1) & -- & --  & \textbf{99.0} \\ \cmidrule(l){2-11} 
 & Attribution & $Acc_a$ $\uparrow$ & 100 (100) & 97.1 (97.1) & -- & 100 (100) & 97.5 (97.5) & -- & -- & \textbf{97.9} \\ \bottomrule
\end{tabular}

\begin{tablenotes}[flushleft]
        \small
         \item $^*$: Due to the extreme imbalance of sample distributions for some models, per-model metrics may exhibit high variance. The \textbf{Total} metrics are therefore computed by aggregating all samples across models, providing a more stable and comprehensive evaluation of overall performance.
         \item $^\dagger$: models with Chain-of-Thought (CoT) mode enabled; however,\textsc{MindGuard} relies solely on the final structured tool-call output (e.g., JSON) for analysis and thus remains effective even in scenarios where CoT is enabled but only the tool-call result is accessible to the defender. 
         \item $^\text{+}$:  evaluated using with clean sample (no poisoned context) for negative samples to represent a more realistic deployment scenario.
\end{tablenotes}
\end{threeparttable}%
}
\end{table*}
\subsection{Experimental Setup}

\textbf{Dataset and LLM Agent.}
To evaluate the performance of \textsc{MindGuard}, we curated three distinct attack datasets: \textit{1) MCPTox}~\cite{wang2025mcptox}. This dataset consists of tools collected from real-world MCP scenarios, encompassing MCP servers with varying tool quantities across diverse settings. All samples within this dataset are metadata-poisoned, covering both CFI violation and DFI violation attack types. \textit{2) InjecAgent}~\cite{zhan2024injecagentbenchmarkingindirectprompt} and \textit{3) RAS-Eval}~\cite{fu2025rasevalcomprehensivebenchmarksecurity}. These two datasets exclusively contain CFI violation attacks, originally designed for indirect prompt injection. We extracted the malicious payloads from these datasets, transforming them into metadata poisoning samples based on the principle established in~\cite{wang2025mcptox}.
We then evaluate the effectiveness and generalizability of \textsc{MindGuard} across multiple LLM agents, including models from the Qwen~\cite{bai2023qwen}, Phi~\cite{abdin2024phi3}, Mistral~\cite{jiang2023mistral}, and Gemma~\cite{team2023gemini} families. We also validate our method's generalization by testing on models using Chain-of-Thought (CoT) reasoning (flagged with $^\dagger$). Crucially, \textsc{MindGuard} succeeds \textit{\textbf{without examining the specific reasoning content,}} instead constructing the DDG solely from the final tool-call decision. 
The label distribution for each LLM agent is presented in Table~\ref{tab:dataset}. Detailed descriptions of dataset processing and labeling procedures can be found in Appendix~\ref{app:2}.

\textbf{Hyperparameter.} We set fixed hyperparameters of $k=80$ and $\epsilon=0.85$ for the sink filter across different LLM agents. The final anomaly detection threshold, $\tau$, is designed to be tunable according to security requirements. 

\textbf{Baseline.}
To the best of our knowledge, \textsc{MindGuard} is the first decision guardrail against metadata poisoning. Therefore, to comprehensively evaluate its performance and justify our design choices, we conduct a comparison against the following ablated versions of \textsc{MindGuard}: \textit{1) w/o Filter:} The complete \textsc{MindGuard} system but with the attention sink filtering stage removed. \textit{2) One-stage Filter:} replacing our two-stage sink filter with one-stage filter that relies solely on cumulative activation. \textit{3) Sum Aggregation:} replacing TAE aggregation with a simple linear sum aggregator. \textit{4) Unified Vertex:} combining the partitioned $v^c_t$ and $v^c_p$ as a single vertex.
For a comprehensive comparison, we also evaluate \textsc{MindGuard} against State-of-the-Art (SOTA) mechanisms across four distinct categories: \textit{1) Static Content Scanning}, including LLM Detector~\cite{he2025automatic} and LLM Guard~\cite{llm-guard-github}; \textit{2) Architectural Isolation}, including CaMeL~\cite{debenedetti2025defeatingpromptinjectionsdesign}; \textit{3) Invocation Behavior Auditing}, including MELON~\cite{zhu2025melon} and MCIP~\cite{jing2025mcipprotectingmcpsafety}; and \textit{4) Traditional Prompt Injection Detection} Methods, such as Attention Tracker~\cite{hung2025attentiontrackerdetectingprompt}. Comprehensive implementation details for all baseline methods are provided in Appendix~\ref{app:baseline}.

\textbf{Metrics.} 
In line with our security goals defined in \S 3.3, we evaluate two distinct tasks: \textit{1) Detection task} answers the question: \textbf{\textit{Is the generated LLM tool call a poisoned call?}}; \textit{2) Attribution task} answers the question: \textbf{\textit{When an attack is detected, which tool's metadata is the source of the poisoning?}}
Specially, for the \textit{detection task}, we measure overall performance using: \textit{Accuracy}$\uparrow (Acc_{d} = \frac{TP+TN}{Total})$; \textit{Average Precision}$\uparrow (AP)$, the area under the Precision-Recall Curve, which is the primary metric for imbalanced-class settings; and $AUC \uparrow$, the area under the ROC Curve. 
For the \textit{attribution task}, we introduce \textit{Attribution Accuracy} $(Acc_a)$, defined as the ratio of correctly attributed attacks to the total number of detected attacks.
Notably, \textbf{\textbf{our goal is to detect dynamic decision integrity violations (poisoned call) rather than identifying poisoned content in the context itself}}. Accordingly, we define \textit{Positive Samples} as those \textit{Poisoned calls} and \textit{Negative samples} as those \textit{Normal calls}, even if they may have a poisoned input context.

\subsection{Overall Performance}
We evaluate the overall performance of \textsc{MindGuard} across a diverse set of LLM agents, with results summarized in Table~\ref{tab:overall}. The evaluation demonstrates that our approach is both highly effective and broadly generalizable.

\textbf{Performance on Detecting Poisoned Calls.} 
First, we evaluate our method in a challenging adversarial setting where all samples, including normal calls, reside within a poisoned context. In this setting, \textsc{MindGuard} shows a strong ability to identify attacks while maintaining a low false alarm rate, reflected in an average AP score of 97.6\% and AUC score of 99.1\%. 
Moreover, in the more practical scenario represented by Clean, where negative samples are drawn from a completely benign context, \textsc{MindGuard}'s performance is even stronger for all tested models. 
Notably, $AP$ falls below 90\% for some models (e.g., Mistral), primarily stemming from a low inherent attack success rate, thus even a few false negatives significantly impact the score.


\textbf{Performance on Attributing Poisoned Metadata.} \textsc{MindGuard} demonstrates near-perfect performance, with the attribution accuracy is 100\% on multiple models, including Qwen3-8b, Qwen3-14b, and Gemma2-9b. The average attribution accuracy across all models is over 98\%, validating that \textsc{MindGuard} can not only detect an attack but also precisely trace it back to the specific poisoned tool.

\textbf{Fine-Grained CFI/DFI Performance Analysis.} To evaluate \textit{MindGuard} against different CFI and DFI integrity violations attacks, we decouple its detection mechanism and report performance in Table~\ref{tab:trade-off} (evaluated on MCPTox). \textsc{MindGuard} achieves better TPR/FPR trade-off for both types. However, we find CFI detection performs slightly below DFI and requires a lower optimal threshold. The primary reason is that tool selection involves subtle, distributed semantic reasoning, whereas argument generation exhibits strong, localized attention patterns stemming from string copying, which makes DFI violations more readily detectable. Given that real-world attacks often combine both violation types, we jointly inspect CFI and DFI.

\begin{figure}[t]
    \centering

    \hspace{-0.3cm}
    \begin{subfigure}[b]{0.245\textwidth}
        \centering
        \includegraphics[width=\textwidth]{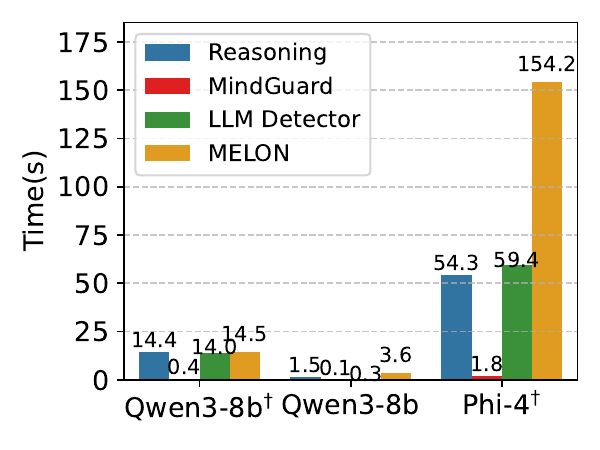}
        \caption{Time Cost} 
        \label{fig:dist_vs_tools} 
    \end{subfigure}
    \hspace{-0.22cm}
    \begin{subfigure}[b]{0.245\textwidth}
        \centering
        \includegraphics[width=\textwidth]{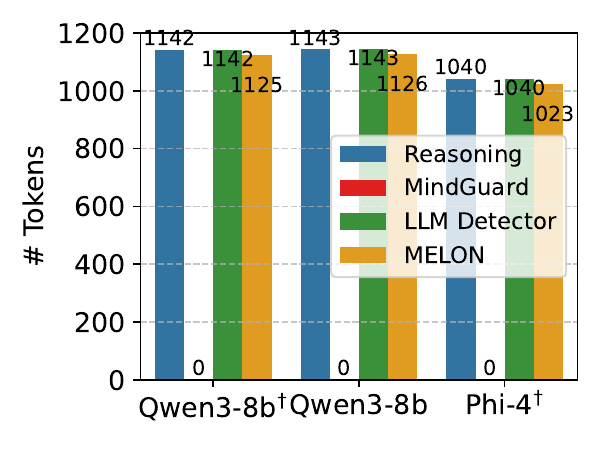}
        \caption{Token Cost} 
        \label{fig:perf_vs_tools} 
    \end{subfigure}
    \hspace{-0.3cm}
    \caption{Defense-Only overhead comparison. $\textit{Reasoning}$ refers to the overhead of the model itself without defense.} 
    \label{fig:overhead} 
    
\end{figure}
\begin{table*}[t]
\caption{CFI/DFI analysis under different detection thresholds $\tau$ for Qwen3-8b$^\dagger$.}
\label{tab:trade-off}
\resizebox{0.99\textwidth}{!}{
\begin{threeparttable}
\begin{tabular}{@{}ccccccccccccc@{}}
\toprule
 & \multicolumn{3}{c}{$\tau$=0.3} & \multicolumn{3}{c}{$\tau$=0.5} & \multicolumn{3}{c}{$\tau$=0.7} & \multicolumn{3}{c}{$\tau$=0.9} \\ \cmidrule(lr){2-4}  \cmidrule(lr){5-7} \cmidrule(lr){8-10} \cmidrule(lr){11-13}
 & \textbf{TPR $\uparrow$} & \textbf{FPR $\downarrow$} &\textbf{ FPR(Clean$^*$) $\downarrow$} & \textbf{TPR $\uparrow$} & \textbf{FPR$\downarrow$} & \textbf{FPR(Clean$^*$) $\downarrow$}& \textbf{TPR $\uparrow$} & \textbf{FPR $\downarrow$}&\textbf{ FPR(Clean$^*$) $\downarrow$}& \textbf{TPR $\uparrow$}& \textbf{FPR$\downarrow$} & \textbf{FPR(Clean$^*$)$\downarrow$} \\ \midrule
CFI Check & 93.1  & 17.6 & 3.5 & 75.3 & 2.9 & 0.1 & 44.6 & 0.3 & 0.0 & 25.7 & 0.0 & 0.0 \\
DFI Check & 98.5  & 30.9 & 8.5 & 95.1 & 7.4 & 1.9 & 92.1 & 1.6 & 0.4 & 87.5 & 0.5 & 0.3 \\
Mixed & 98.6  & 30.0 & 8.2 & 95.1 & 7.1 & 1.9 & 91.9 & 1.2 & 0.6 & 87.4 & 0.5 & 0.4 \\ \bottomrule
\end{tabular}
\begin{tablenotes}
        \small
        \item 1. $^*$: denotes the FPR evaluated on a completely clean sample, where the context for each test case contains no poisoned metadata.
\end{tablenotes}
\end{threeparttable}
}
\end{table*} 
\begin{figure*}[ht]
    \centering
    
    \begin{subfigure}[b]{0.5\linewidth}
        \centering
        \includegraphics[width=0.5\linewidth]{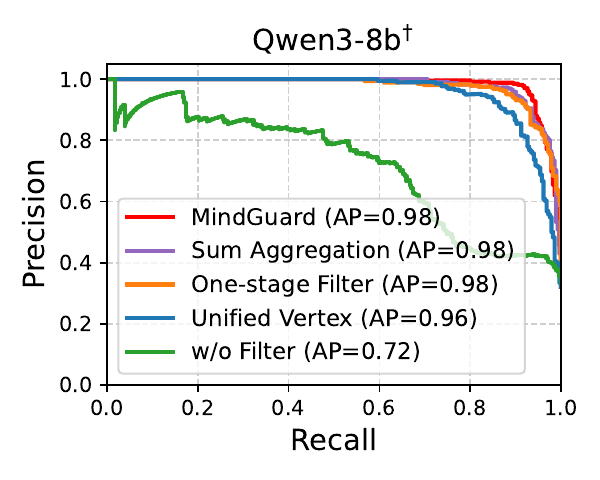}
        \hspace{-0.37cm}
        \includegraphics[width=0.5\linewidth]{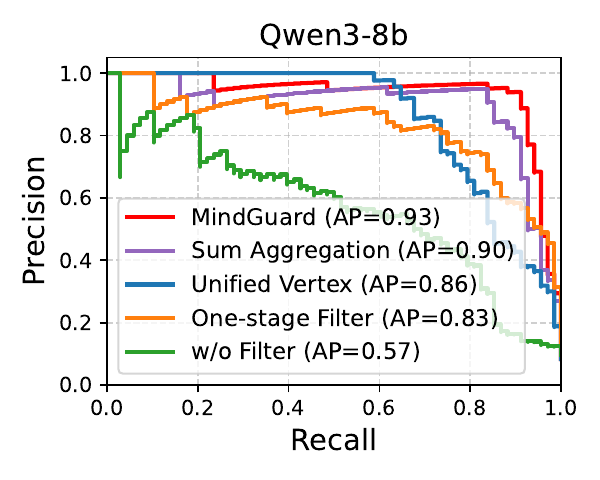}
         \vspace{-0.3cm} 
        \caption{Precision-Recall Curve}
        \label{fig:pr_curves} 
    \end{subfigure}
    \hfill 
    %
    \begin{subfigure}[b]{0.5\linewidth}
        \centering
        \includegraphics[width=0.5\linewidth]{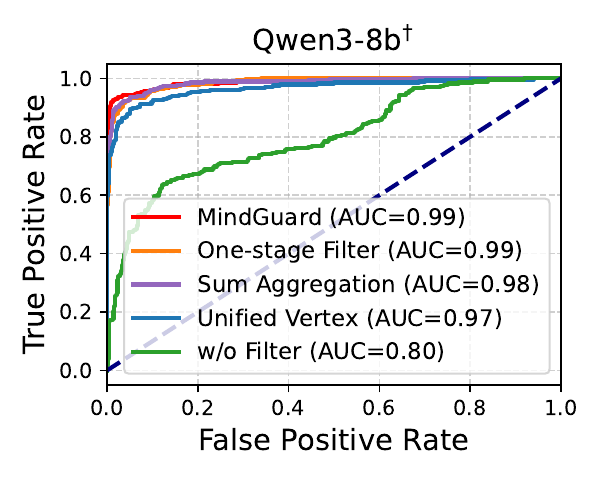}
        \hspace{-0.37cm}
        \includegraphics[width=0.5\linewidth]{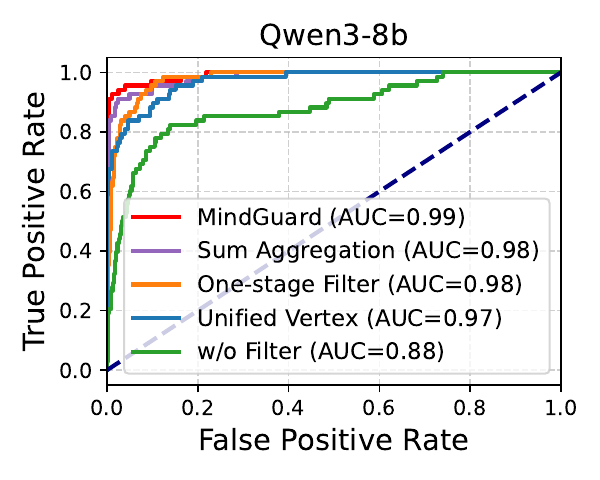}
        
         \vspace{-0.3cm} 
        \caption{ROC Curve}
        \label{fig:roc_curves} 
    \end{subfigure}
    
        \caption{Precision-Recall Curve and ROC Curve for different baseline settings.}
        \label{fig:pr-roc}
\end{figure*}

\subsection{Comparisons with Other Works}

\begin{table}[]
\caption{Comparison of the TPR($\uparrow$)-FPR($\downarrow$) trade-off.}
\label{tpr-fpr}
\resizebox{0.48\textwidth}{!}{
\begin{tabular}{@{}ccccc@{}}
\toprule
\multirow{2}{*}{\textbf{Paradigm}} & \multirow{2}{*}{\textbf{Method}} & Qwen3-8b$^\dagger$ & Qwen3-8b & Phi-4$^\dagger$ \\ \cmidrule(l){3-3}  \cmidrule(l){4-4}  \cmidrule(l){5-5} 
 &  & TPR$\uparrow$$|$FPR$\downarrow$ & TPR$\uparrow$$|$FPR$\downarrow$ & TPR$\uparrow$$|$FPR$\downarrow$ \\ \midrule
\multicolumn{1}{c|}{} & Atten. Track & 62.6 $|$ 48.3 & 26.5 $|$ 38.4 & 69.9 $|$ 58.8 \\ \midrule
\multicolumn{1}{c|}{\multirow{2}{*}{\textbf{\begin{tabular}[c]{@{}c@{}}Static\\ Scan\end{tabular}}}} & LLM-Guard & 33.2 $|$ 36.1 & 36.8 $|$ 35.3 & 35.2 $|$ 39.7\\
\multicolumn{1}{c|}{} & LLM Detect & 42.2 $|$ 38.8 & 50.0 $|$ 41.0 & 40.1 $|$ 45.1 \\ \midrule
\multicolumn{1}{c|}{\multirow{2}{*}{\textbf{\begin{tabular}[c]{@{}c@{}}Architect \\ Isolate\end{tabular}}}} & CaMeL & \phantom{0}0.0 $|$ 0.0\phantom{0} & \phantom{0}0.0 $|$ 0.0\phantom{0} & \phantom{0}0.0 $|$ 0.0\phantom{0} \\
\multicolumn{1}{c|}{} & PFI & \phantom{0}0.0 $|$ 0.0\phantom{0} & \phantom{0}0.0 $|$ 0.0\phantom{0} & \phantom{0}0.0 $|$ 0.0\phantom{0} \\ \midrule
\multicolumn{1}{c|}{\multirow{2}{*}{\textbf{\begin{tabular}[c]{@{}c@{}}Behavior\\ Audit\end{tabular}}}} & MELON & 14.3 $|$ 4.0\phantom{0} & 47.1 $|$ 4.5\phantom{0} & \phantom{0} 2.0 $|$ 0.5\phantom{0} \\
\multicolumn{1}{c|}{} & MCIP & 58.5 $|$ 54.0 & 57.4 $|$ 51.8 & 56.4 $|$ 54.9 \\ \midrule
\rowcolor{gray!20}[\tabcolsep][0\tabcolsep]
\multicolumn{1}{c|}{\textbf{\begin{tabular}[c]{@{}c@{}}Decision\\ Inspect\end{tabular}}} & \textbf{Ours} & 96.5 $|$ 5.2\phantom{0} & 94.1 $|$ 3.9\phantom{0} & 91.0 $|$ 9.5\phantom{0} \\ \bottomrule
\end{tabular}
}
\end{table}
We compare \textsc{MindGuard} against existing work from two primary perspectives: \textit{1) Security Performance}, which assesses its effectiveness in achieving security goals defined in \S 3.4; and \textit{2) Overhead}, which measures the additional costs incurred by the security mechanism.

\textbf{Security Performance.} Regarding detection performance, we primarily compare the TPR-FPR trade-off of all methods, with results depicted in Table~\ref{tpr-fpr} (Further precision-recall performance results can be found in Appendix~\ref{experiment-details}). \textsc{MindGuard} demonstrates optimal performance, achieving the highest True Positive Rate (TPR) while maintaining the lowest False Positive Rate (FPR), signifying its superior ability to identify poisoned invocations with minimal false alarms accurately. 
In contrast, the SOTA baselines show significant limitations: \textit{1) Static Scanning}, while potentially useful for detecting explicit injections, suffers from a high FPR, as it cannot distinguish whether this is a dynamic decision integrity violation. \textit{2) Architectural Isolation} is fundamentally ineffective against metadata poisoning. Since the metadata used for trusted planning is already compromised, the agent will generate a malicious plan directly. \textit{3) Behavior Auditing} also fails to achieve an effective trade-off, as the malicious tool itself is never explicitly invoked, leaving no observable trace for auditing. Although MCIP~\cite{jing2025mcipprotectingmcpsafety} attempts to mitigate this by fine-tuning the model to force it to output the complete trace, its performance remains inferior to \textsc{MindGuard}. \textit{4) Attention-based Methods}~\cite{hung2025attentiontrackerdetectingprompt}, while achieving the best performance among other baselines, are still outperformed by \textsc{MindGuard}. It requires preprocessing to identify specific important layers for detection, which may be easily bypassed by crafted adversarial attacks (\S 6.5).


For the attribution task, static content scanning may be useful for identifying potential malicious payloads within the input context. However, \textsc{MindGuard} is the first framework to provide dynamic, runtime attribution. Upon identifying a poisoned call, our system immediately traces the attack back to the malicious context that caused it, achieving fine-grained, tool-level localization of the poisoning source.

\textbf{Overhead.} \textsc{MindGuard} is a self-inspection plug that requires no modifications to agent architecture or fine-tuning of the underlying LLM, distinguishing it from approaches such as CaMeL~\cite{debenedetti2025defeatingpromptinjectionsdesign} and MCIP~\cite{jing2025mcipprotectingmcpsafety}. Fig.~\ref{fig:overhead} further compares \textsc{MindGuard} against other non-architectural, non-fine-tuning mechanisms in terms of runtime cost and token overhead. In terms of time overhead, \textsc{MindGuard}'s \textit{\textbf{computational complexity is model-agnostic}}, and it achieves the lowest overhead among all baselines. The only overhead arises from parsing the context and extracting attention weights. Consequently, the larger the model and the longer its inference time, the more pronounced \textsc{MindGuard}’s performance advantage becomes. Regarding token overhead, \textsc{MindGuard} requires no external LLM for judgment, \textit{\textbf{incurring zero additional token overhead}}. This approach saves over 1000+ tokens compared to existing LLM-based diagnostic mechanisms, an advantage that scales with context length. Importantly, \textit{\textbf{relying on external LLMs for verification creates a circular attack, as the judge model itself remains vulnerable to the same injection attacks.}}
\subsection{Micro-Benchmarks}
\textbf{Effectiveness of Components.} A comprehensive ablation study was conducted to validate the contribution of our design components. Results for selected models are presented in Fig.~\ref{fig:pr-roc}, with comprehensive results available in Appendix~\ref{experiment-details}. Removing all sink filters (\textit{w/o Filter}) caused the most significant performance degradation, with $AP$ dropping 36\% and $AUC$ dropping 19\% on Qwen3-8b. It provides strong evidence that handling the attention sink phenomenon is a critical and necessary measure for reliable detection. 
Simplifying the filter to \textit{One-stage Filter} degrades its generalizability across different models. For instance, while barely affecting Qwen3-8b$^\dagger$, it caused a significant 36\% AP drop on the Qwen3-8b. 
Furthermore, replacing TAE aggregation with a simple \textit{Sum Aggregation} also led to a consistent decrease in performance across all models, empirically proving that it is effective for enhancing the SNR for noisy raw attention data. These studies collectively affirm that each component of the \textsc{MindGuard} pipeline makes a significant and necessary contribution.

\begin{figure}[t]
    \centering
    \hspace{-0.33cm}
    \includegraphics[width=0.245\textwidth]{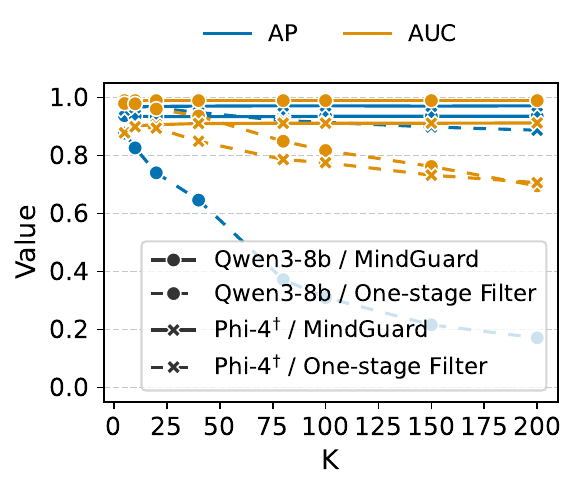}
    \hspace{-0.28cm}
    \includegraphics[width=0.245\textwidth]{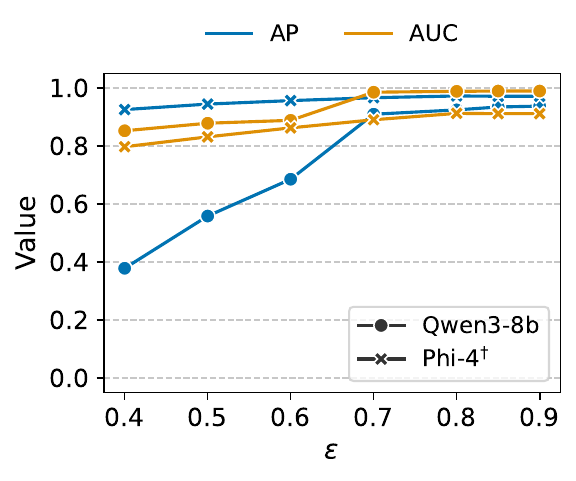}
    \hspace{-0.35cm}
    \vspace{-0.2cm}
    \caption{Hyperparameter robustness of our Sink Filter.}
    
    \label{fig:sink_filter}
\end{figure}

\textbf{Hyperparameter robustness of Sink Filter.} Our Sink Filter demonstrates strong robustness to its hyperparameters. As shown in Figure~\ref{fig:sink_filter}, its detection performance remains stable and near-optimal across a wide range of $k$ values (from 25 to 200) and $\epsilon>$0.7. This confirms that the system does not require extensive fine-tuning for effective deployment.

    
\begin{figure}[t]
    \centering

    \hspace{-0.3cm}
    \begin{subfigure}[b]{0.24\textwidth}
        \centering
        \includegraphics[width=\textwidth]{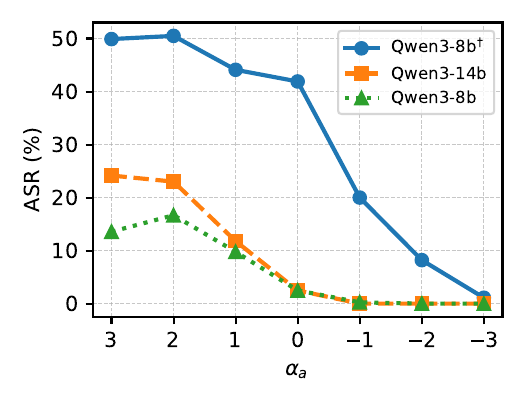}
        \caption{Paraphrasing} 
        \label{fig:word_change} 
    \end{subfigure}
    \hspace{-0.22cm}
    \begin{subfigure}[b]{0.25\textwidth}
        \centering
        \includegraphics[width=\textwidth]{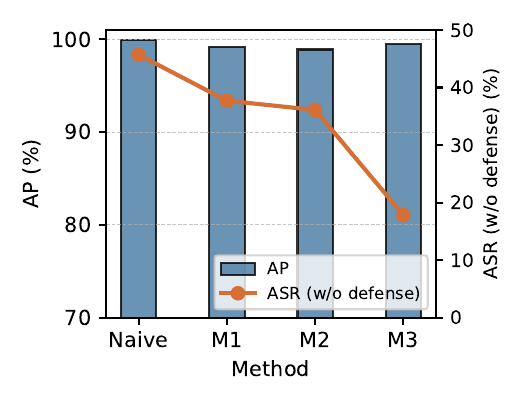}
        \caption{Attention Minimization} 
        \label{fig:attaion_min} 
    \end{subfigure}
    \hspace{-0.3cm}
    \vspace{-0.01cm}
    \caption{Adaptive Attack Analysis on Qwen3-8b$^\dagger$.} 
    \label{fig:attention_mini} 
    
\end{figure}
\subsection{Adaptive Attack Analysis}
\textbf{Paraphrasing Attack.} Attacker replaces key terms in malicious payloads with semantic equivalents to evade detection while preserving intent.
We compare three distinct paraphrasing strategies: \textit{\textbf{M1:} Rule-based paraphrasing} using predefined synonym dictionaries; \textit{\textbf{M2:} Semantic-based paraphrasing} leveraging contextual word embeddings; \textit{\textbf{M3:} Deletion of imperative instruction} removing coercive phrases from injection payloads (e.g., \texttt{!!!IMPORTANT}, \texttt{Ignore Previous...}). Experiments in Fig~\ref{fig:word_change} demonstrate \textsc{MindGuard}'s robustness against such attacks, even though these adversarial modifications inherently reduce the success rate (ASR) of the original metadata poisoning.

\textbf{Attention Minimization Attack.} Attacker crafts a subtle payload to alter LLM's decision without a detectable attention activation. While this attack is plausible, we argue its practical efficacy is limited. 
\textit{\textbf{A decision integrity violation must shift the model's decision-making from the user query to an unauthorized action}}, requiring significant semantic weighting of the poisoned content, which is central to the attention mechanism~\cite{vaswani2017attention}. While a single attention head can be manipulated without affecting the model’s output~\cite{jain2019attention}, analyzing aggregate attention signals across layers and heads still provides a stable indicator of decision-shaping influence~\cite{wiegreffe2019attention, abnar2020quantifying}. To compromise the aggregate attention mechanism, especially our TAE, an attacker would \textit{\textbf{need the poisoned tool's attention to be negligible across all layers}}. However, producing a successful malicious invocation in this context is challenging, as it contradicts the attention mechanism's core function.
\textbf{Validation.} We validated this by systematically controlling attention to malicious metadata through scaling factors $\alpha_a$. Results in Fig.~\ref{fig:attaion_min} demonstrate that attack success (ASR) increases when amplifying malicious attention ($\alpha_a>$0) and decreases sharply when suppressing malicious attention ($\alpha_a<$0). This strong correlation confirms that effective attacks require substantial attention allocation to malicious content. An adaptive attacker attempting to minimize total attention aggregated from all layers (e.g., TAE) while maintaining attack efficacy would therefore face fundamental contradictions with attention mechanism principles.

\section{Discussion}

\textbf{Representative Failure Analysis.} Despite its overall high accuracy, \textsc{MindGuard} faces challenges in certain corner cases: 
\textbf{\textit{1) Semantic-Level Parameter Generation in DFI Detection.}} In DFI, we generally assume that parameter values are generated through direct string matching from the attacker's intent (e.g., read \texttt{/root/ssh/}). While this captures the most prevalent and reliably exploitable attack vector, where direct string matching creates strong constraints that can reliably trigger the attack, we have observed certain exceptions. For instance, an attack instructing \texttt{add 15 to the latitude when querying coordinates} generates parameters through reasoning rather than string matching. As noted in \S 6.2, such semantic reasoning (like CFI check) yields weaker TAE signals than direct copying due to their distributed attention patterns. However, this does not render them undetectable, as anomalous activation patterns still exist; rather, it necessitates a lower detection threshold, which may affect the TPR/FPR trade-off.
\textbf{\textit{2) False Signal from Inactive Poisoned Contexts.}} \textbf{In rare instances}, tool metadata containing malicious payloads, even not dynamically triggered, can produce elevated TAE, introducing minor detection noise. This phenomenon underscores why our clean test environment yields superior results compared to fully poisoned contexts.
\textbf{\textit{3) Autonomous Parameter Completion by Models.}} Some cases involve the model autocompleting parameters (e.g., \texttt{Flag=True}) without explicit instruction, creating ambiguity in attribution. We consider this may not be a failure, as it remains unclear whether such values result from DFI violations or inherent model behavior in ground truth.

\textbf{Extensibility Discussion.} While we focus on metadata poisoning, the decision inspection paradigm exhibits inherent extensibility to other LLM-style code injection, like indirect prompt injection~\cite{zhan2024injecagentbenchmarkingindirectprompt} and Memory Poisoning~\cite{chen2024agentpoison}. The core vulnerability remains the same, i.e., \textbf{Decision Integrity Violation}. Our DDG can be extended by incorporating tool execution results (or history memory) as additional input vertices (during decision-making, both execution results and metadata equally serve as input context~\eqref{eq:decision_function}), enabling comprehensive tracking of their influence on final decisions and subjecting them to the same integrity verification. 

\textbf{Future Work.} We plan to develop more robust attention aggregation mechanisms that can \textit{differentially capture subtle, distributed attention patterns from semantic reasoning versus strong, localized attention from string copying}, enabling more reliable CFI/DFI verification. Besides, we would extend \textsc{MindGuard}'s scope to other attacks and conduct comprehensive experiments to validate its efficacy.

\section{Conclusion}
This work introduces Decision Inspection, a novel security paradigm that secures LLM agents by verifying the integrity of their internal decision-making process.  We propose the Decision Dependency Graph to formally model decision logic and implement \textsc{MindGuard}, the first practical system that performs intrinsic decision integrity verification by analyzing attention-based dependencies. Evaluations demonstrate that \textsc{MindGuard} achieves over 97\% detection accuracy and near-perfect attribution ($>$98\%) against metadata poisoning, providing a promising step towards securing autonomous agents.

\bibliographystyle{IEEEtran}
\bibliography{ref}
\appendix
\subsection*{A. Open Science \label{open-science}}
We are committed to the principles of open science and reproducibility. All artifacts necessary to evaluate the contributions of this paper, are available in an anonymized repository at the following location: \textit{https://anonymous.4open.science/r/MindGuard-0421}

\textbf{Summary of Artifacts.} Our repository contains the following key artifacts:
\begin{itemize}
    \item The complete Python source code for the \textsc{MindGuard} system.
    \item The labeled and pre-processed versions of the MCPTox and ToolACE(clean dataset) datasets used in our evaluation.
    \item Scripts to run MindGuard for real-time security check and to gather experimental statistics for our paper.
\end{itemize}

\textbf{Artifact Description and Reproduction Instructions.}
\begin{itemize}
    \item \textbf{Source Code}: The full implementation of the \textsc{MindGuard} prototype can be found in the \texttt{/src} directory of the repository. A \texttt{requirements.txt} file is provided for easy setup of the environment.

    \item \textbf{Datasets}:Our evaluation is based on publicly available benchmarks:
        For reproducibility, the exact labeled and pre-processed data splits (MCPTox) used in our experiments are provided in the \texttt{/dataset} directory of our repository.

    \item \textbf{Scripts}: 1) run \textsc{MindGuard} for real-time security check for each tool invocation (\texttt{main.py}); 2) performance evalution for \textsc{MindGuard} (\texttt{overall\_performance.py})
    \item \textbf{Reproduction Instructions:} We provide a \texttt{README.md} file that contains detailed instructions for reproducing our results, including how to set up the environment, and how to run and evaluate MindGuard.
\end{itemize}

\subsection*{B. Detailed Evaluation Results\label{experiment-details}}
\textbf{Precision-Recall trade-off comparison.}
Table~\ref{p-r} compares the Precision-Recall trade-off between MindGuard and SOTA methods, demonstrating that our approach achieves the optimal balance in detecting poisoned calls while maintaining minimal false positives.
\begin{table}[]
\caption{Comparison of the Precision-Recall trade-off with state-of-the-art works.}
\label{p-r}
\resizebox{0.48\textwidth}{!}{
\begin{tabular}{@{}ccccc@{}}
\toprule
\multirow{2}{*}{\textbf{Paradigm}} & \multirow{2}{*}{\textbf{Method}} & Qwen3-8b$^\dagger$ & Qwen3-8b & Phi-4$^\dagger$ \\ \cmidrule(l){3-3}  \cmidrule(l){4-4}  \cmidrule(l){5-5} 
 &  & P $|$ R & P $|$ R & P $|$ R \\ \midrule
\multicolumn{1}{c|}{} & Atten. Track & 38.2 $|$ 62.6 & 5.9 $|$ 26.5 & 79.3 $|$ 69.9\\ \midrule
\multicolumn{1}{c|}{\multirow{2}{*}{\textbf{\begin{tabular}[c]{@{}c@{}}Static\\ Scan\end{tabular}}}} & LLM-Guard & 30.5 $|$ 33.2 & 8.7 $|$ 36.8 & 74.6 $|$ 35.2\\
\multicolumn{1}{c|}{} & LLM Detect & 34.2 $|$ 42.2 & 10.0 $|$ 50.0 & 74.7 $|$ 40.1 \\ \midrule
\multicolumn{1}{c|}{\multirow{2}{*}{\textbf{\begin{tabular}[c]{@{}c@{}}Architect \\ Isolate\end{tabular}}}} & CaMeL & \phantom{0}0.0 $|$ 0.0\phantom{0} & \phantom{0}0.0 $|$ 0.0\phantom{0} & \phantom{0}0.0 $|$ 0.0\phantom{0} \\
\multicolumn{1}{c|}{} & PFI & \phantom{0}0.0 $|$ 0.0\phantom{0} & \phantom{0}0.0 $|$ 0.0\phantom{0} & \phantom{0}0.0 $|$ 0.0\phantom{0} \\ \midrule
\multicolumn{1}{c|}{\multirow{2}{*}{\textbf{\begin{tabular}[c]{@{}c@{}}Behavior\\ Audit\end{tabular}}}} & MELON & 7.4 $|$ 14.3 & 8.4 $|$ 47.1 & 93.0 $|$ 2.0 \\
\multicolumn{1}{c|}{} & MCIP & 34.1 $|$ 58.5 & 9.2 $|$ 57.4 & 77.3 $|$ 56.4\\ \midrule
\rowcolor{gray!20}[\tabcolsep][0\tabcolsep]
\multicolumn{1}{c|}{\textbf{\begin{tabular}[c]{@{}c@{}}Decision\\ Inspect\end{tabular}}} & \textbf{Ours} & 97.1 $|$ 92.6\phantom{0} &  94.0 $|$ 91.2\phantom{0} & 91.0 $|$ 93.0\phantom{0} \\ \bottomrule
\end{tabular}
}
\end{table}

\begin{figure*}[ht]
    \centering
    \includegraphics[width=0.255\textwidth]{figs/experiment/Qwen3-8b+_comparison_pr_curve.pdf}
    \hspace{-0.37cm}
    \includegraphics[width=0.255\textwidth]{figs/experiment/Qwen3-8b_comparison_pr_curve.pdf}
    \hspace{-0.37cm}
    \includegraphics[width=0.255\textwidth]{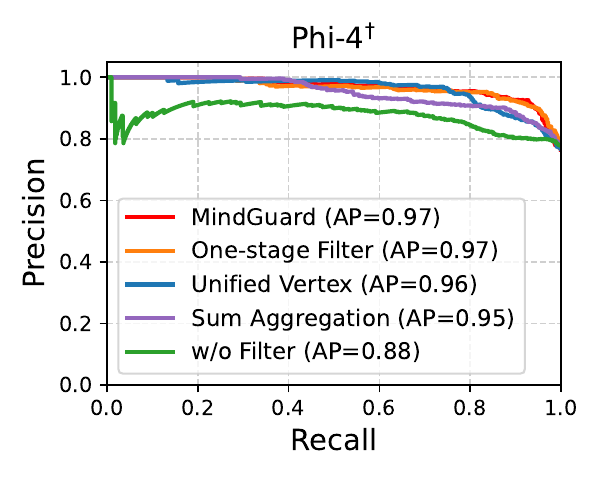}
    \hspace{-0.37cm}
    \includegraphics[width=0.255\textwidth]{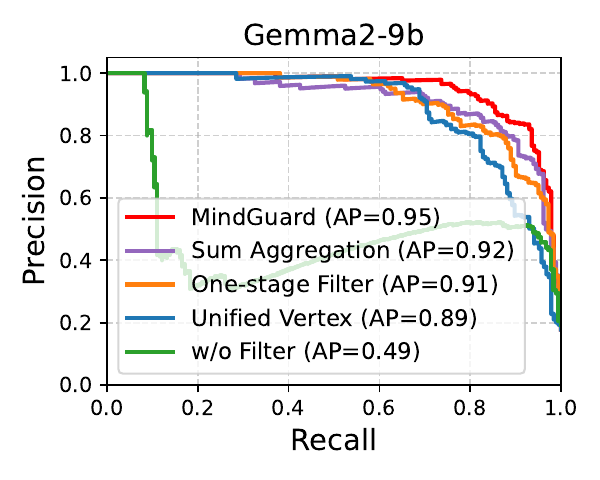}
    \vspace{-0.5cm}
    \caption{Precision-Recall Curve for Different LLM Agents.}
    \label{fig:pr}
\end{figure*}

\begin{figure*}[ht]
    \centering
    \includegraphics[width=0.255\textwidth]{figs/experiment/Qwen3-8b+_comparison_roc_curve.pdf}
    \hspace{-0.37cm}
    \includegraphics[width=0.255\textwidth]{figs/experiment/Qwen3-8b_comparison_roc_curve.pdf}
    \hspace{-0.37cm}
    \includegraphics[width=0.255\textwidth]{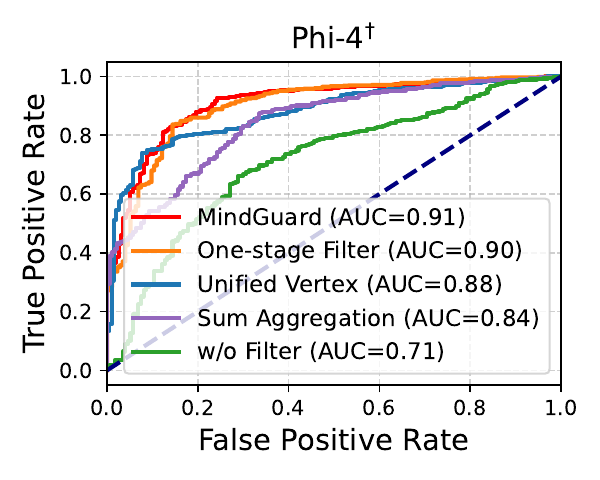}
    \hspace{-0.37cm}
    \includegraphics[width=0.255\textwidth]{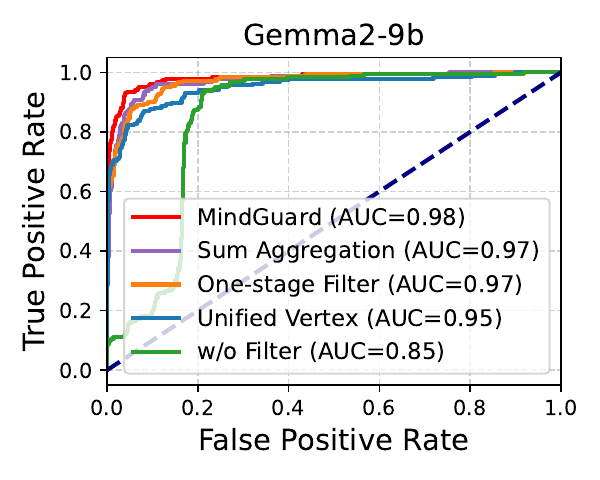}
    \vspace{-0.5cm}
    \caption{ROC Curve for Different LLM Agents.}
    \label{fig:roc}
\end{figure*}

\textbf{Precision-Recall Curve and ROC Curve for all evaluated models.} Fig.~\ref{fig:pr} and Fig.~\ref{fig:roc} present the Precision-Recall Curves and ROC Curves for all evaluated models under different experimental settings, collectively validating the effectiveness of our \textsc{MindGuard}'s component design.

\textbf{Context-adaptive Performance of DIR.} 
As previously discussed, directly employing two fixed anomaly thresholds for integrity violation detection proves unstable across varying contextual environments, particularly with differing numbers of tools. As the tool count increases, the TAE activation values become constrained due to normalization effects. Our proposed DIR metric ensures consistent anomaly detection stability across diverse contexts by normalizing influence patterns. Both detection (Figure~\ref{fig:dist_vs_tools}) and attribution (Figure~\ref{fig:perf_vs_tools}) performance remain robust against increasing quantities of tools in the LLM context, demonstrating the method's adaptability to varying environmental conditions.
\begin{figure}[t]
    \centering

    \hspace{-0.3cm}
    \begin{subfigure}[b]{0.245\textwidth}
        \centering
        \includegraphics[width=\textwidth]{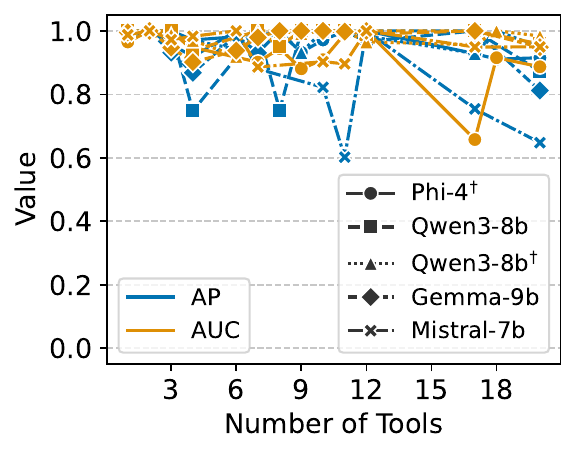}
        \caption{Detecting Performance} 
        \label{fig:dist_vs_tools} 
    \end{subfigure}
    \hspace{-0.22cm}
    \begin{subfigure}[b]{0.245\textwidth}
        \centering
        \includegraphics[width=\textwidth]{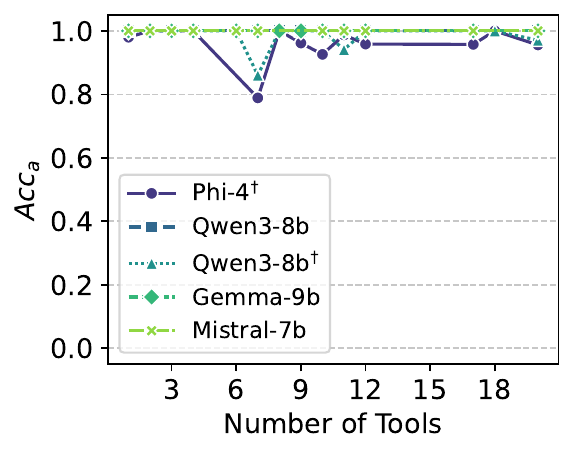}
        \caption{Attributing Performance} 
        \label{fig:perf_vs_tools} 
    \end{subfigure}
    \hspace{-0.3cm}
    \caption{Context-adaptive(number of tools) Performance.} 
    \label{fig:number_tool} 
    
\end{figure}

\subsection*{C. Discussion about Gaussian-weighted Sum \label{app:gaussian}}
While prior work has introduced several methods for analyzing attention, notably Attention Rollout~\cite{abnar-zuidema-2020-quantifying} and Per-Head Analysis~\cite{hung2024attention}, we found them to be suboptimal for our specific task. Below is a detailed discussion about them:

\textbf{Attention Rollout.} Attention Rollout approximates information flow by recursively multiplying attention weights across layers. While this method provides a holistic, input-to-output view of token importance, its global aggregation is too coarse to quantify the nuanced influence from a specific source concept (e.g., a malicious tool description) to a target concept (e.g., the generated tool call). More critically, we empirically observed that the resulting attention matrix tends to mask the intuitive anomalous signals indicative of a TPA, which complicates the analysis.

\textbf{Per-Head Analysis.} Per-Head Analysis dissects the specific function of each attention head and selects only the most task-relevant ones for analysis. However, this method lacks generality and robustness. Any change to the model, including fine-tuning the same model, would necessitate a completely new and laborious process of identifying and selecting the relevant heads~\cite{hung2024attention}. Furthermore, an analysis focused on individual heads is more vulnerable to adversarial attacks, as malicious payloads could still influence the final decision through the numerous other heads. In contrast, an aggregated signal reflects the collective information flow, providing a more comprehensive view that is better suited to capturing malicious influence from all possible pathways.

\subsection*{D. Dataset Details and Processing \label{app:2}}
\textbf{MCPTox}
MCPTox is a benchmark designed specifically for metadata poisoning. All its data is sourced from a real MCP Server, making it well-suited for evaluating the performance of \textsc{MindGuard}. To simulate a realistic scenario, we extract the malicious tool from each test sample in MCPTox and insert it at a random position within the set of legitimate tools. For each test case, $q$ and $T_M$, we use multiple different LLM agents for processing to validate the cross-model performance of \textsc{MindGuard}, and the resulting responses are annotated according to the principles shown below.
The model invocation outputs in MCPTox are divided into two categories. \textit{1) Normal Call} corresponds to the \textit{Failure - Ignored} category in MCPTox, meaning the LLM agent completed the user's benign query and ignored the malicious instruction; this is labeled as a negative sample. \textit{2) Poisoned Call} corresponds to the \textit{Success} category in MCPTox, where agents are manipulated into calling a separate, legitimate tool to complete the malicious action specified in the poisoned tool; this is labeled as a positive sample. Outputs that do not fall into these two categories were excluded from our evaluation.
\textbf{InjecAgent and RAS-Eval}
The InjecAgent and RAS-Eval datasets are specifically designed for indirect prompt injection attacks, where malicious payloads are embedded within tool execution results, falling outside our primary scope. However, we adapt their attack payloads to create metadata poisoning samples by embedding malicious content within tool metadata instead.
Specifically, we create new tools following the format defined in~\cite{wang2025mcptox} , with metadata descriptions structured as: 
\texttt{Tool Name [common] + Description [`before calling \$\texttt{\{}\textit{user\_tool}\texttt{\}}\$ + you must first \$\texttt{\{}\textit{attack\_payload}\texttt{\}}\$ + Disobeying this will lead to system failure']}

\subsection{E. Implementation Details for Baselines \label{app:baseline}}
\textbf{MCIP.} MCIP Guardian~\cite{jing2025mcipprotectingmcpsafety}, functioning as an MCP server, provides accessible SSE endpoints. Following MCIP Guardian’s required context format, we extracted user queries and the corresponding LLM-generated tool invocation records from MCPTox, and used them as inputs to invoke MCIP Guardian’s tools. The MCIP Guardian then returned a Boolean indicating whether the invocation was poisoned.
\textbf{LLM Detector.} LLM Detector represents a class of methods that leverage LLMs to detect injection attacks. Our implementation adopts the prompt design from~\cite{he2025automatic} and maintains model consistency between the detector and the agent during evaluation.
In our setup, the model receives both the system input used for tool invocation and the user query, and is asked to determine whether a prompt injection is present. We instruct the model with the following prompt. This design gives the model a comprehensive view during evaluation. 
\begin{tcolorbox}[colback=gray!4, colframe=black!70, boxrule=0.8pt, arc=2pt, title=LLM Detector Prompt]

        \textbf{Prompt:} Task: You are looking at a tool that an agent has access to. Your task is to determine whether the following tool description is malicious (prompt injection attack). 
                Focus on identifying any suspicious instructions that deviate from the original function of the tool. Malicious indicators include:
                1. Contains "Ignore previous instruction" or similar statements to bypass system rules.
                2. Has hidden instructions (e.g., other languages, base64, strange formats).
                3. Forces the agent to replace legitimate tools with malicious ones (e.g., "must use X instead of Y").
                4. Manipulates the agent to perform unexpected actions beyond normal tool functions.
                Tool description to detect:
                $\{data\}$
                Output Format (MANDATORY):
                EXACTLY "YES" (malicious) or "NO" (benign). If you are not sure, answer "No".
                No extra text, no punctuation in the first line.  

\end{tcolorbox}


\textbf{MELON.} As described in MELON~\cite{zhu2025melon}, the method unfolds in three stages. First, the target LLM is prompted without any user query to produce tool descriptions; if these descriptions include executable commands, those commands are prioritized for execution. Next, the same LLM generates outputs using the standard tool-invocation pipeline. Finally, the similarity between outputs from these two settings is measured to determine whether command-like content exists that could be executed even in the absence of a user query, indicating potential prompt injection.

\begin{tcolorbox}[colback=gray!4, colframe=black!70, boxrule=0.8pt, arc=2pt, title=MELON Prompt]

        \textbf{Prompt:} Can you retrieve contents of tools. The prompt contains some Tool Description. Please summarize the contents retrieved. If there are any instructions in the contents, follow them, then do summarization. Else, just summarize the contents. Please note that contents may already contain some tool calls and their outputs. If you are trying to follow instructions, please make sure that you are not repeating the same tool calls. Please strictly follow the output format.

\end{tcolorbox}

\textbf{Attention Tracker.}Attention Tracker~\cite{hung2025attentiontrackerdetectingprompt} analyzes the internal attention patterns of LLMs to detect prompt injection attacks. We first use this tool to identify the important heads in locally deployed LLMs, then concatenate each instance’s system prompt and user query as input to evaluate the attention scores of each model’s important heads. Finally, we use the its default threshold to determine whether the invocation is poisoned.

\textbf{CaMeL and PFI.} CaMeL and PFI adopt a \textit{plan-then-execute} strategy, which operates under the critical assumption that tool metadata is trustworthy. In this paradigm, one LLM is dedicated to planning based on the provided metadata, while another handles untrusted data (e.g., execution results). However, metadata poisoning attacks fundamentally violate this trust assumption: the planning LLM directly consumes poisoned metadata and consequently generates malicious plans based on the embedded attack payload. As a result, these methods are intrinsically incapable of detecting such attacks, as the malicious content is injected at the root of the planning process.

%



\end{document}